\documentclass[review]{elsarticle}

\usepackage{lineno,hyperref}
\modulolinenumbers[5]

\journal{Journal of Artificial Intelligence}

\usepackage{enumitem}
\usepackage{graphicx}
\usepackage{algorithm}
\usepackage{algpseudocode}
\usepackage{subcaption}

\usepackage{titlesec}

\usepackage{amssymb,amsmath,amsthm,amsfonts}

\usepackage[T1]{fontenc}

\newtheorem{theorem}{Theorem}[section]
\newtheorem{lemma}[theorem]{Property}
\newtheorem{lemmaz}[theorem]{Lemma}
\newtheorem{prop}{Property}[section]

\newtheorem{definition}{Definition}

\newenvironment{hproof}{%
  \proof}{\endproof}

\newenvironment{zproof}{%
  \proof}{\endproof}

\usepackage{tabularx,environ,amsmath,amssymb}

\makeatletter
\newcommand{\problemtitle}[1]{\gdef\@problemtitle{#1}}% Store problem title
\newcommand{\probleminput}[1]{\gdef\@probleminput{#1}}% Store problem input
\newcommand{\problemquestion}[1]{\gdef\@problemquestion{#1}}% Store problem question
\NewEnviron{problem}{
  \problemtitle{}\probleminput{}\problemquestion{}% Default input is empty
  \BODY% Parse input
  \par\addvspace{.5\baselineskip}
  \noindent
  \begin{tabularx}{\textwidth}{@{\hspace{\parindent}} l X c}
    \multicolumn{2}{@{\hspace{\parindent}}l}{\@problemtitle} \\% Title
    \textbf{Instance:} & \@probleminput \\% Input
    \textbf{Question:} & \@problemquestion% Question
  \end{tabularx}
  \par\addvspace{.5\baselineskip}
}

\usepackage[margin=1.0in]{geometry}

\begin{document}

\begin{frontmatter}

\title{The Computational Complexity of Angry Birds}

\author{Matthew Stephenson}
\address{Department of Data Science and Knowledge Engineering, Maastricht University, Maastricht, The Netherlands}
\ead{matthew.stephenson@maastrichtuniversity.nl}
\author{Jochen Renz, Xiaoyu Ge}
\address{Research School of Computer Science, Australian National University, Canberra, Australia}

\begin{abstract}
The physics-based simulation game Angry Birds has been heavily researched by the AI community over the past five years, and has been the subject of a popular AI competition that is currently held annually as part of a leading AI conference. Developing intelligent agents that can play this game effectively has been an incredibly complex and challenging problem for traditional AI techniques to solve, even though the game is simple enough that any human player could learn and master it within a short time. In this paper we analyse how hard the problem really is, presenting several proofs for the computational complexity of Angry Birds. By using a combination of several gadgets within this game's environment, we are able to demonstrate that the decision problem of solving general levels for different versions of Angry Birds is either NP-hard, PSPACE-hard, PSPACE-complete or EXPTIME-hard. Proof of NP-hardness is by reduction from 3-SAT, whilst proof of PSPACE-hardness is by reduction from True Quantified Boolean Formula (TQBF). Proof of EXPTIME-hardness is by reduction from G2, a known EXPTIME-complete problem similar to that used for many previous games such as Chess, Go and Checkers. To the best of our knowledge, this is the first time that a single-player game has been proven EXPTIME-hard. This is achieved by using stochastic game engine dynamics to effectively model the real world, or in our case the physics simulator, as the opponent against which we are playing. These proofs can also be extended to other physics-based games with similar mechanics.
\end{abstract}


\end{frontmatter}

\section{Introduction}
The computational complexity of different video games has been the subject of much investigation over the past decade. However, this has mostly been carried out on traditional style platformers \cite{ori1,ori4} or primitive puzzle games \cite{ori8,ori5}. In this paper, we analyse the complexity of playing different variants of the video game Angry Birds, which is a sophisticated physics-based puzzle game with a semi-realistic and controlled environment \cite{web}. The objective of each level in this game is to hit a number of pre-defined targets (pigs) with a certain number of shots (birds) taken from a fixed location (slingshot), often utilising or avoiding blocks and other game elements to achieve this. An example of an Angry Birds level is shown in Figure 1. Angry Birds is a game of great interest to the wider AI research community due to the complex planning and physical reasoning required to solve its levels, similar to that of many real-world problems. It has also been used in the AIBIRDS competition \cite{ABcomp} which tasks entrants with developing agents to solve unknown Angry Birds levels and aims to promote the integration of different AI areas \cite{extra5}. Many of the previous agents that have participated in this competition employ a variety of AI techniques, including qualitative reasoning \cite{cite3}, internal simulation analysis \cite{cite5,cite4}, logic programming \cite{cite10}, heuristics \cite{cite11}, Bayesian inferences \cite{cite7,cite6}, and structural analysis \cite{cite8}. Despite many different attempts over the past five years the problem is still largely unsolved, with AI approaches far from human-level performance. 

\begin{figure}
  \centering
  \includegraphics[width=0.75\linewidth]{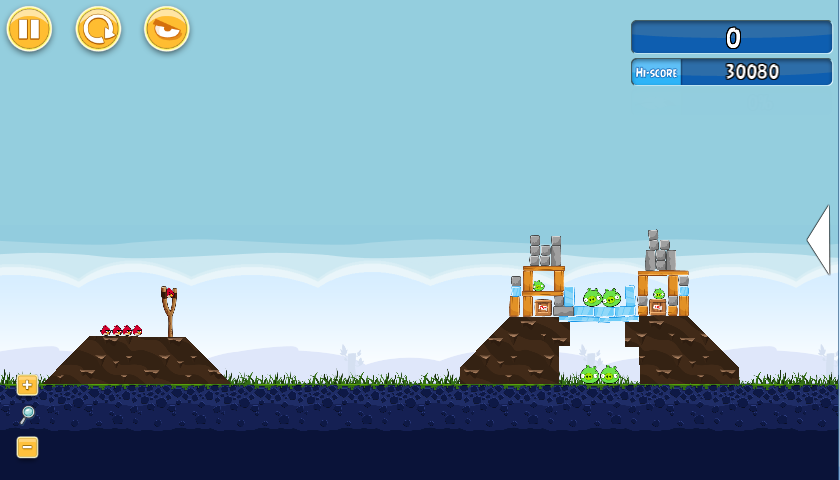}
  \caption{Screenshot of a level for the Angry Birds game.}
\end{figure}

Video games have been the subject of much prior research on computational complexity, with many papers proving specific games to be either NP-hard or PSPACE-complete. Examples of past proofs for NP-hardness include games such as Pac-Man \cite{ori5}, Lemmings \cite{ori3}, Portal \cite{ori7}, Candy Crush \cite{ori17}, Bejeweled \cite{ori18}, Minesweeper \cite{ori19}, Tetris \cite{ori22}, and multiple classic Nintendo games \cite{ori1}. Proofs of PSPACE-completeness have also been described for games such as Mario Bros. \cite{ori9}, Doom \cite{ori5}, Pok\'{e}mon \cite{ori1}, Rush Hour \cite{extra3}, Mario Kart \cite{mariokart} and Prince of Persia \cite{ori4}. Interestingly, the video game Braid has been proven to be PSPACE-hard \cite{extra4} but not PSPACE-complete. However, none of these video games have yet been proven EXPTIME-hard. Proofs of EXPTIME-hardness have previously been demonstrated for several traditional two-player board games, including Chess \cite{exp3}, Checkers \cite{exp2} and the Japanese version of Go \cite{exp4}. As far as we are aware, no single-player video game without a traditional opponent has ever been proven EXPTIME-hard before now.

Complexity proofs have also been presented for many different block pushing puzzle games, including Sokoban \cite{ori14}, Bloxorz \cite{ori12} and multiple varieties of PushPush \cite{ori2,ori13,ori15}. These proofs have been used to advance our understanding of motion planning models due to their real-world similarities \cite{ori11}. It is therefore important that the computational complexity of physics-based games is investigated further, as playing video games such as Angry Birds has much in common with other real-world AI and robotics problems \cite{new1}. A physics-based environment is very different from that of traditional games as the attributes and parameters of various objects are often imprecise or unknown, meaning that it is very difficult to accurately predict the outcome of any action taken. Angry Birds also differs from many previously investigated games in terms of its control scheme, as the player always makes their shots from the same location within each level (slingshot position) and can only vary the speed and angle at which each bird travels from it. This heavily reduces the amount of control that the player has over the bird's movement, with the game's physics engine being used to determine the outcome of shots after they are made.

The remainder of this paper is organised as follows: Section 2 formally defines the Angry Birds game, as well as the different variants of it that will be used within our proofs; Section 3 describes the designs and workings of several gates that will be used in later proofs; Sections 4 - 7 present proofs that particular variants of Angry Birds are either PSPACE-complete, PSPACE-hard, NP-hard or EXPTIME-hard respectively; Section 8 provides some suggestions and examples of how the presented proofs could be extended to other games with similar mechanics; Section 9 concludes this work and proposes future possibilities.

\section{Angry Birds Game Definition}
Angry Birds is a popular physics-based puzzle game in which the objective is to kill all the pigs within a 2D level space using a set number of birds. Each level has a predefined size and any game element that moves outside of its boundaries is destroyed. The area below the level space is comprised of solid ground that cannot be moved or changed in any way, although other elements can be placed on or bounced off of it. Players make their shots sequentially and in a predefined order, with all birds being fired from the location of the slingshot. The player can alter the speed (up to a set maximum) and angle with which these birds are fired from the slingshot but cannot alter the bird's flight trajectory after doing so, except in the case of some special bird types with secondary effects that can be activated by the player. Once a bird has been fired, it is removed from the level after not moving for a certain period of time. The level space can also contain many other game elements, such as blocks, static terrain, explosives, etc. All game elements have a positive fixed mass, friction, dimensions and shape (based on their type), and no element may overlap any other. Birds that have yet to be fired are the only exception to this rule and may overlap other elements within the level space (i.e. birds do not interact physically with other game elements until fired from the slingshot; they are simply visible within the level for visual effect). The level itself also has a fixed gravitational force that always acts downwards.  If two objects collide they will typically bounce off each other or one of the objects will break. Calculations done with regard to object movement and resolving collisions are simulated using a simplified physics engine based on Newtonian mechanics. The exact mathematics and physical rules of how the engine works are not provided as this would be incredibly long and tedious. Instead all proofs presented in this paper are done at a high level, allowing the concepts and ideas to be easily extended to other similar games or problems. All level designs presented in subsequent sections have taken the specific physics of the engine into consideration and can be demonstrated to work within the original Angry Birds game environment.

The description of an Angry Birds level can be formalised as $Level = (L_{x}, L_{y}, slingshot, birds, pigs, other)$.
\begin{itemize}  
\item $L_{x}$ is the width of the level in pixels.
\item $L_{y}$ is the height of the level in pixels.
\item $slingshot$ is the pixel coordinates $(x,y)$ from which the player makes their shots.
\item $birds$ is a list containing the number $(N_{b})$, type and order of the birds available.
\item $pigs$ is a list containing the type, angle and pixel coordinates $(x,y)$ of all the pigs.
\item $other$ is a list containing the type, angle and pixel coordinates $(x,y)$ of all other game elements; including blocks, static terrain and other miscellaneous objects not considered for our presented proofs.
\end{itemize}

The top left corner of a level is given the coordinates $(0,0)$ and all other coordinates use this as a reference point. The width and height of a level must be specified as non-negative integer values, and all pixel coordinates must be defined as integers within the level space. 
All numerical values are assumed to be stored in binary, meaning that the size of a given level description is logarithmic with respect to the values inside of it.
The precision with which the angle of a pig or other game element within the level description can be defined is set to some arbitrary value (e.g. 0.01 degrees) as the rotation of objects is not important for the proofs presented in this paper. The type of a bird, pig or other game element is defined using a fixed length word (e.g. ``red'' or ``small''). How the number of birds $(N_{b})$ is defined greatly impacts the complexity of the game, with further details on this point described in Section 2.1.
There is also a finite sized list which contains all the possible types of birds, pigs and other game elements, as well as their properties (e.g. mass, friction, size, etc.). This list is fixed in size and so is not relevant to the complexity of the game.

One important point that must be addressed is how the properties of certain game elements (position, angle, speed, etc.) are represented within the game engine. 
Whilst the initial location of each game element is defined using integer values (pixel coordinates), when the game is being played it is highly likely that the location of an object could be much more precise (i.e. sub-pixel values). For our proofs we assume that the current state of a level, including the current properties of all game elements within it, can always be stored in a polynomial number of bits.

A strategy $(S)$ for solving a given level description consists of a sequence of ordered shots ($A_{1},A_{2},...,A_{N_{b}}$). Each shot $(A_{i})$ consists of both a pixel coordinates $(x,y)$ within the level space (release point), which determines the speed $(v_{b})$ and angle $(a_{b})$ with which each of the available birds is fired, as well as a tap time for activating each bird's secondary effect (ability) if it has one. For our presented proofs we do not use any bird abilities, meaning that a particular shot $A_{i}$ can be defined using just a release point $(x,y)$. A level is won/solved once all pigs have been killed, and is lost/unsolved if there are any pigs left once all birds have been used. 

While the speed with which a bird can be fired is bounded, and therefore can only be determined to a set level of precision, the angle of a shot is a rational value that is determined by the release point given. The tap time for activating a bird's ability must occur before the bird collides with another game element or moves out of bounds. Therefore, the precision with which shots can be specified, as well as the number of bits required to define a shot and the number of distinct shots possible, is polynomial in the size of the level (i.e. the size of the level dictates the number of possible release points/shot angles and tap times, which in turn determines the number of distinct shots possible), and is exponential relative to the size of the level description (as all numerical values are specified in binary). This means that the number of possible distinct shots that a player can make increases as the size of the level increases (i.e. no fixed arbitrary precision on possible shot angles), but this number is always bounded by the size of the level ($L_{x} \times L_{y}$). 

The decision problem we are considering in this paper can be formalised as:
\begin{problem}
 \problemtitle{\textbf{Angry Birds Formal Decision Problem}}
  \probleminput{Angry Birds level description $(Level)$.}
  \problemquestion{Is there a strategy $S$ that always results in all $pigs$ being killed?}
\end{problem}

This is the same problem that is faced by both level designers and play testers for this game.

For the proofs described in this paper the following game elements are required:
\begin{itemize}  
\item Red Birds: These are the most basic bird type within the game and possess no special abilities. Once the player has determined the speed and angle with which to fire this bird it follows a trajectory determined by both this and the gravity of the level, which the player cannot subsequently affect. This bird has no secondary effect so a tap time is not needed.
\item Small Pigs: These are the most basic pig type within the game and are killed once they are hit by either a bird or block.
\item Unbreakable Blocks: These are blocks that do not break if they are hit but instead react in a semi-realistic physical way, moving and rotating if forces are applied to them. They are represented in this paper by blocks made of stone.
\item Static Terrain: This is simply a set area of the level that cannot move or be destroyed. Static terrain is also not affected by gravity, meaning that it can be suspended in the air without anything else holding it up. It is represented in this paper by plain, untextured, brown areas. The ground at the bottom of the level space also behaves in the same way as static terrain.
\end{itemize}

For our proofs, we assume that the size of a level is not bounded by the game engine and that the player's next shot only occurs once all game elements are stationary. We also assume that the physics calculations performed by the game engine are not impacted or affected as the size of the level increases (i.e. no glitches or other simulation errors) and that there is no arbitrary fixed precision with regard to the angles that shots can have (i.e. the number of distinct shots possible always increases and decreases based on the size of the level). As the exact physics engine parameters used for Angry Birds are not currently available for analysis, all assumptions made about the game and its underlying properties are determined through careful observation.

\subsection{Game Variants}
While an Angry Birds level that is created using the above description can be shown to be at least NP-hard, by making additional specifications on the type of physics engine used or how a level is described, we can increase its complexity further. Deciding whether a particular version of Angry Birds is NP-hard, PSPACE-hard, PSPACE-complete or EXPTIME-hard is based on a combination of two factors.

\textbf{Number of Birds:}
The first factor is whether the number of birds that the player has is polynomial or exponential relative to the size of the level description. In practical terms this means, does the type and order of each bird have to specified individually (i.e. an explicit list of all bird types, e.g. [red, blue, black, red, yellow]) or can the number of birds simply be stated if all birds are the same type (i.e. [red, 5] rather than [red, red, red, red, red])? If this abbreviated version of $birds$ is valid within the level description then the player can potentially have an exponential number of birds, otherwise only a polynomial number of birds is possible.

\textbf{Probabilistic Model:}
The second factor is whether the physics engine used by the game is deterministic or stochastic. A game engine that is deterministic will always base its output only on the player's input, and so the outcome of any action can be calculated in advance.
However, if the game engine is stochastic in nature then physical interactions between game elements may be influenced slightly by randomly generated values. This randomness within the engine is used to simulate the effects of unknown variables in the real world. Specific real-world properties such as air movement (wind), temperature fluctuations, differences in the gravitational field, object vibrations, etc., might affect the outcome of a physical action. These effects are usually not modelled and add some stochasticity to the outcome of physical actions. For Angry Birds, the source of this stochasticity comes from a random amount of noise that is included when collisions occur within the game's physics-engine, causing the object(s) involved in the collision to move slightly differently each time. This means that even if the same collision occurs multiple times for the exact same level state, the outcome may not always be the same. These changes are typically not very large, often only affecting the outcome very slightly within a pre-defined range of options. While the player might know the different outcomes that an action could have, they may not know exactly which one will occur until after said action is performed.
Please note that for the sake of our proofs we consider a game containing elements with pseudorandom behaviour/physics to still be deterministic, as long as the random seed used to define them can be encoded in a polynomial number of bits (i.e. not truly stochastic) \cite{ori1}.

\begin{table}
\begin{center}
\begin{tabular}{|p{3.8cm}|p{4.0cm}|p{2.6cm}|p{3.8cm}|}
    \hline
    \multicolumn{3}{|c|}{\textbf{Game Version}} & \\ \hline
    \textbf{Number of Birds} & \textbf{Probabilistic Model} & \textbf{Acronym} & \textbf{Complexity} \\ \hline
Polynomial & Deterministic & ABPD & NP-hard \\ \hline
Exponential & Deterministic & ABED & PSPACE-complete \\ \hline
Polynomial & Stochastic & ABPS & PSPACE-hard \\ \hline
Exponential & Stochastic & ABES & EXPTIME-hard \\ \hline
  \end{tabular}
\end{center}
\captionof{table}{Complexity results summary.}
\end{table}

Table 1 shows how altering these two factors within the Angry Birds game affects its complexity. For each of our subsequent complexity proofs, we will assume that we are using the appropriate version of Angry Birds as defined by this table. These different game versions will be abbreviated as ABPD for our NP-hard variant, ABED for our PSPACE-complete variant, ABPS for our PSPACE-hard variant, and ABES for our EXPTIME-hard variant.

\section{Gates}
Before presenting our complexity proofs we will first define three different ``gates'' as well as a Crossover, that help dictate the outcomes of shots taken by the player. The design and behaviour of these gates is described here so that they can be easily referred to in later sections. Depending on the specific physics parameters of the environment and objects used, the exact values used to define each gate's design may vary. However, a gate that works for certain velocities and gravitational forces can always be created. The design and parameters of these gates have been fine-tuned for the Angry Birds game engine to prevent elements within them from moving in unintended ways, but could easily be generalised to different game environments.

\subsection{Selector Gate}
The Selector gate implementation for Angry Birds is shown in Figure 2. 
The Selector gate can exist in one of two states, ``select-left'' or ``select-right'', and essentially mimics the behaviour of a 2-output demultiplexer. A summary of the Selector gate behaviour is shown in Table 2.

\begin{figure}[t]
  \centering
	\begin{subfigure}{0.406\textwidth}
  \centering
	\includegraphics[width=\textwidth]{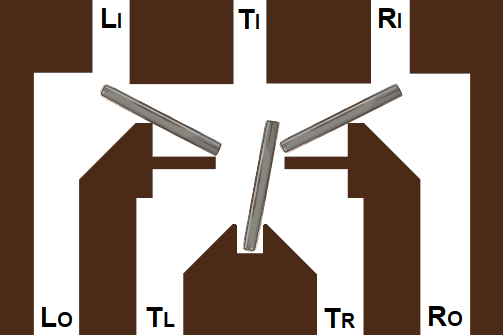}
    \caption{}
	    \end{subfigure}
~
    \begin{subfigure}{0.4\textwidth}
  \centering
        \includegraphics[width=\textwidth]{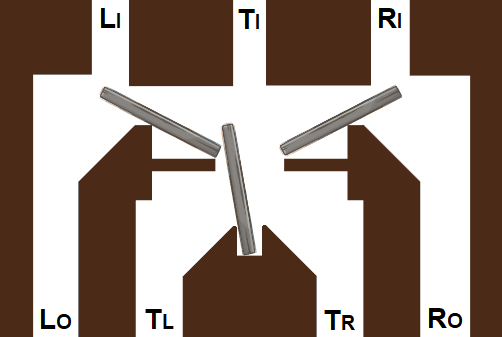}
        \caption{}
    \end{subfigure}
\caption{Models of the Selector gate (a) in the ``select-left'' position and (b) in the ``select-right'' position.}
\end{figure}

\begin{table}[t]
\small
\begin{center}
\begin{tabular}{|p{2cm}|p{2cm}|p{3.2cm}|p{2cm}|p{3.2cm}|}
    \multicolumn{5}{c}{\textbf{Selector gate}} \\ \hline
    & \multicolumn{4}{c|}{\textbf{Current gate position}} \\ \hline
    & \multicolumn{2}{c|}{select-left} & \multicolumn{2}{c|}{select-right} \\ \hline
    \textbf{Entrance} & \textbf{Exit} & \textbf{Next gate position} & \textbf{Exit} & \textbf{Next gate position} \\ \hline
    \textbf{$T_{I}$} & $T_{L}$ & select-left &  $T_{R}$ & select-right \\  \hline
    \textbf{$L_{I}$} & $L_{O}$ & select-left & $L_{O}$ & select-left \\ \hline
    \textbf{$R_{I}$} & $R_{O}$ & select-right & $R_{O}$ & select-right \\ \hline
  \end{tabular}
\end{center}
\captionof{table}{Selector gate summary, shows exits and next gate positions for given entrances and current gate positions.}
\end{table}

\begin{prop}
A bird which enters a Selector gate at $T_{I}$ will exit the Selector gate at $T_{L}$, if and only if the Selector gate is in the select-left position. Otherwise the bird will exit out of $T_{R}$.
\end{prop}

\begin{prop}
A bird which enters a Selector gate at $L_{I}$ will exit the Selector gate at $L_{O}$ and set the Selector gate to the select-left position.
\end{prop}

\begin{prop}
A bird which enters a Selector gate at $R_{I}$ will exit the Selector gate at $R_{O}$ and set the Selector gate to the select-right position.
\end{prop}

\subsection{Automatically Unsetting Transfer Gate}
The Automatically Unsetting Transfer Gate (AUT gate) implementation for Angry Birds is shown in Figure 3.
The AUT gate can exist in one of two states, ``select-left'' or ``select-right''. A summary of the AUT gate behaviour is shown in Table 3.

\begin{prop}
A bird which enters an AUT gate at $T_{I}$ will exit the AUT gate at $T_{L}$ and set the AUT gate to the select-right position, if and only if the AUT gate is in the select-left position. Otherwise the bird will exit out of $T_{R}$ and not change the AUT gate's position.
\end{prop}

\begin{prop}
A bird which enters an AUT gate at $L_{I}$ will exit the AUT gate at $L_{O}$ and set the AUT gate to the select-left position.
\end{prop}

\begin{figure}[t]
  \centering
	\begin{subfigure}{0.3\textwidth}
  \centering
	\includegraphics[width=\textwidth]{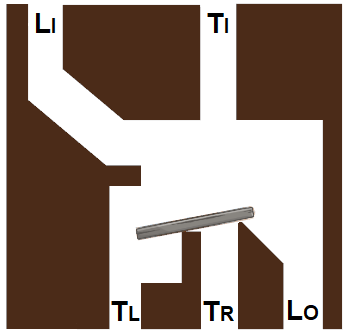}
    \caption{}
	    \end{subfigure}
~
    \begin{subfigure}{0.3\textwidth}
  \centering
        \includegraphics[width=\textwidth]{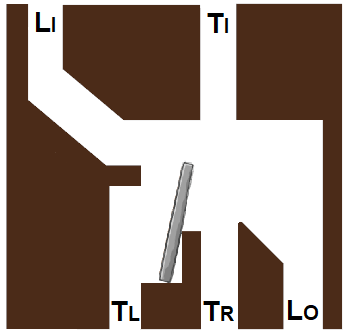}
        \caption{}
    \end{subfigure}
\caption{Models of the AUT gate (a) in the ``select-left'' position and (b) in the ``select-right'' position.}
\end{figure}

\begin{table}[t]
\small
\begin{center}
\begin{tabular}{|p{2cm}|p{2cm}|p{3.2cm}|p{2cm}|p{3.2cm}|}
    \multicolumn{5}{c}{\textbf{AUT gate}} \\ \hline
    & \multicolumn{4}{c|}{\textbf{Current gate position}} \\ \hline
    & \multicolumn{2}{c|}{select-left} & \multicolumn{2}{c|}{select-right} \\ \hline
    \textbf{Entrance} & \textbf{Exit} & \textbf{Next gate position} & \textbf{Exit} & \textbf{Next gate position} \\ \hline
    \textbf{$T_{I}$} & $T_{L}$ & select-right &  $T_{R}$ & select-right \\  \hline
    \textbf{$L_{I}$} & $L_{O}$ & select-left & $L_{O}$ & select-left \\ \hline
  \end{tabular}
\end{center}
\captionof{table}{AUT gate summary, shows exits and next gate positions for given entrances and current gate positions.}
\end{table}

\subsection{Random Gate}
The Random gate implementation for Angry Birds is shown in Figure 4. 
The Random gate can only be used in variants of Angry Birds with a stochastic game engine (ABPS and ABES), and essentially mimics the behaviour of a random binary splitter.

\begin{prop}
A bird which enters a Random gate at point $T$ has a non-zero probability of exiting at point $L$ ($P(L)>0$) and a non-zero probability of exiting at point $R$ ($P(R)>0$). 
\end{prop}

\begin{hproof}
When a bird enters a Random gate at $T$, it will hit the tip of the point. When this happens the physics engine will use randomly generated values to slightly alter the physics of the impact, with three possible outcomes: the bird falls down the left tunnel (exit at $L$), the bird falls down the right tunnel (exit at $R$), the bird remains on the point and falls neither left nor right (does not exit the gate). Property 3.8 is true if the probability for each of the first two outcomes occurring is greater than zero, which is the case for the stochastic Angry Birds game environment.
\end{hproof}

\begin{figure}[t]
\centering
\begin{minipage}{.5\linewidth}
  \centering
  \includegraphics[width=0.65\linewidth]{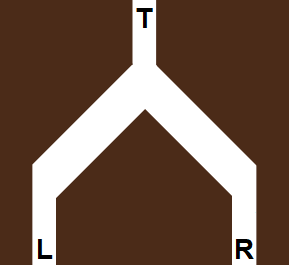}
\captionsetup{justification=centering}
  \captionof{figure}{Model of the Random gate.}
  \label{fig:test2}
\end{minipage}%
\begin{minipage}{.5\linewidth}
  \centering
  \includegraphics[width=.6\linewidth]{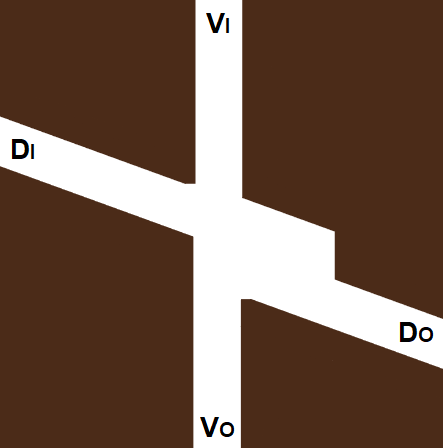}
\captionsetup{justification=centering}
  \captionof{figure}{Model of the Crossover.}
  \label{fig:test1}
\end{minipage}
\end{figure}

\subsection{Crossover}
The Crossover implementation for Angry Birds is shown in Figure 5.

\begin{prop}
A bird which enters a Crossover at $D_{I}$ will exit the Crossover at $D_{O}$.
\end{prop}

\begin{prop}
A bird which enters a Crossover at $V_{I}$ will exit the Crossover at $V_{O}$.
\end{prop}

\begin{figure}[t]
  \centering
  \includegraphics[width=0.8\linewidth]{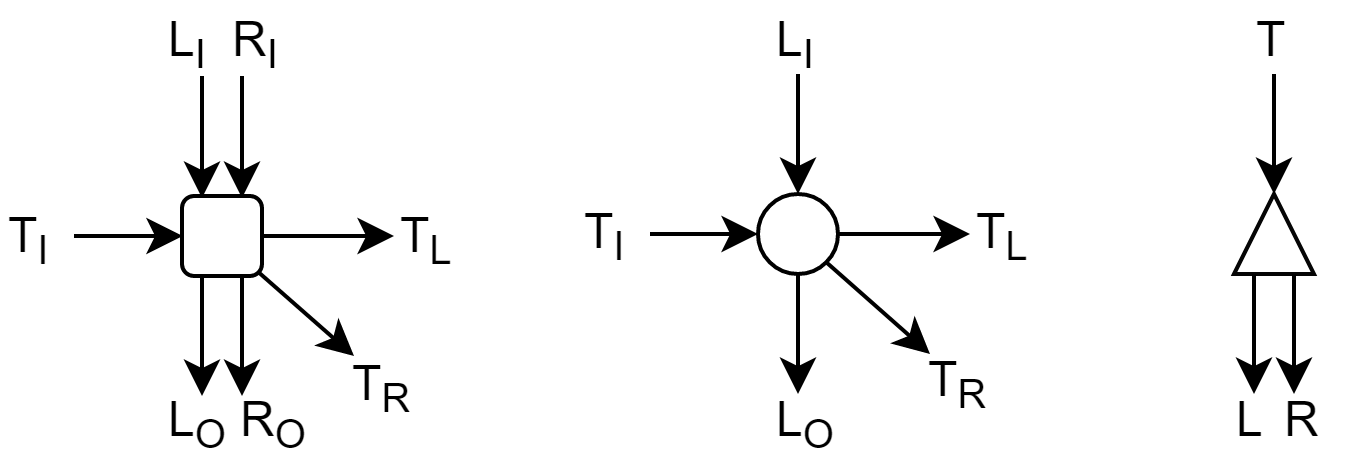}
  \caption{Selector gate (left), AUT gate (middle) and Random gate (right) compact representations, used in our subsequent proof diagrams.}
\end{figure}

\subsection{Gate Representation}

For the diagrams presented in the following proofs we will use a more compact way of representing gates, see Figure 6. Squares represent Selector gates, circles represent AUT gates, and triangles represent Random gates, with the location of the arrows representing the entries to and exits from each gate. Arrows leading from the exit of one gate to the entrance of another, represent tunnels that can be used to connect multiple gates together. A bird will travel along this tunnel, provided that the start of the tunnel is not below the end (bird is essentially falling down the tunnel). If a particular arrow is not given for a specific gate, then that entry or exit is not used (blocked off with static terrain). Any bird that attempts to leave through an exit that is blocked off will be trapped inside the gate, with the bird subsequently disappearing after a short period of time. Crossovers do not have a compact representation, and are instead used to deal with any intersecting tunnels between gates. Note that even though the exact entry and exit locations on the compact gate representations do not match those on the actual gate models/designs, additional tunnels and Crossovers can be easily used to adjust the entry and exit locations for each gate. Therefore an Angry-Birds reduction can be represented by an equivalent "circuit diagram".

\subsection{Terminology}

Selector gates with $T_{R}$ blocked off can be thought of as being very similar to that of a ``door'' mechanism used in several previous video game complexity proofs \cite{ori1,ori5,ori6}. For the sake of both intuitive names and consistent terminology with prior work, we define new terms for our Selector and AUT gates. If a gate is in the select-left position then we say that the gate is ``open'', and if the gate is in the select-right position then we say that the gate is ``closed''. If a gate is open then we say that it can be ``traversed'' by firing a bird into $T_{I}$, which will then exit out of $T_{L}$. A gate can be ``opened'' by firing a bird into $L_{I}$ or ``closed'' by firing a bird into $R_{I}$. Entrances $T_{I}$, $L_{I}$ and $R_{I}$ are referred to as the ``traverse'', ``open'' and ``close'' paths respectively. In subsequent proof diagrams that use the compact gate representation shown in Figure 6, Selector or AUT gates that are initially closed (i.e. select-right) will have a single line border while those that are initially open (i.e. select-left) will have a double line border. This terminology only applies to the Selector and AUT gates, not the Random gate.

\section{PSPACE-Completeness of ABED (exponential and deterministic)}
For our proof of PSPACE-hardness, we will reduce from the PSPACE-complete problem TQBF, which consists of determining if a given quantified 3-CNF Boolean formula is ``true''. In order to demonstrate that Angry Birds is PSPACE-hard, it must be possible to construct a level that represents any given quantified Boolean formula, which can only be solved if the quantified Boolean formula is true (i.e. the player will be able to kill the pig(s) within the level by making shots with their bird(s), if and only if the quantified Boolean formula that the level was based on is true). We can also extend this proof to PSPACE-completeness if the problem of solving ABED levels is also in PSPACE. 
Due to the length and complexity of our presented proofs, this section will be split into the following sub-sections: Section 4.1 describes a high-level overview of the framework that we will use to prove that solving ABED levels is PSAPCE-hard; Section 4.2 describes how we can create the gadgets for this framework within the ABED environment; Section 4.3 describes a method for constructing this framework within the ABED environment using our designed gadgets; Section 4.4 describes a possible winning strategy for an ABED level based on an example quantified Boolean formula; and Section 4.5 proves that solving ABED levels is also in PSPACE.

\subsection{Framework}
For our proof of PSPACE-hardness by TQBF reduction, we will use a heavily modified version of the general framework described in \cite{ori1,ori5,ori6}. This framework uses a systematic procedure to verify if a quantified Boolean formula is true. This process can be defined in general terms, allowing it to be applied to any game environment (including Angry Birds).

\begin{description}
\item[\textbf{TQBF verification process:}]
\end{description}
\begin{enumerate}  
\item The player initially chooses the value of all existentially quantified variables, and the value of all universally quantified variables is set to positive.
\item Check that all clauses within the quantified Boolean formula are satisfied (if not then cannot proceed). 
\item If all universally quantified variables have a negative value, then the quantified Boolean formula is true (verification process complete).
\item The universal quantifier ($UQ_{R}$) with the smallest scope (rightmost universal quantifier in Boolean formula) that has a positive value for its variable, has the value of its variable set to negative. 
\item The player can change the value of any existentially quantified variables within the scope of $UQ_{R}$, and all universally quantified variables within the scope of $UQ_{R}$ are set to positive.
\item Go to step 2.
\end{enumerate}
As an example, given a quantified Boolean formula with three universally quantified variables (x,y,z) of decreasing scope size, the order in which the universal variables are verified is as follows: (1,1,1) (1,1,0) (1,0,1) (1,0,0) (0,1,1) (0,1,0) (0,0,1) (0,0,0).

This process can be successfully completed if and only if the given quantified Boolean formula is true.

While we will still be using this same TQBF verification process for our proposed Angry Birds proof, the overall design of the framework for applying this procedure will be significantly different from those of previous game examples. 
This is mostly due to the fact that Angry Birds does not have a single controllable ``Avatar'', and thus has no easy way of achieving a sense of ``player traversal''. The general design of our TQBF verification framework for Angry Birds is shown in Figure 7.
This framework can be used to prove that a game is PSPACE-hard by constructing the necessary ``gadgets'' (each box within the general framework diagram). Each of these gadgets serves a distinct purpose and simplifies the complex physics of Angry Birds into more easily manageable sections (for our proofs, each gadget is made up of multiple interconnected gates).
For each existential quantifier in the Boolean formula there is an associated Existential Quantifier (EQ) gadget, for each Clause in the Boolean formula there is an associated Clause gadget, and for each universal quantifier in the Boolean formula there is both an associated Universal Quantifier True (UQ-T) gadget and Universal Quantifier False (UQ-F) gadget. There is also a Finish gadget, which the player must be able to ``pass through'' in order to solve the level.
Figure 7 demonstrates an example arrangement of these gadgets using the quantified Boolean formula $\exists x \forall y \exists z \forall w ((x \vee y \vee w) \wedge (y \vee \neg z \vee \neg w) \wedge (\neg x \vee \neg y \vee z))$ as an example (each variable in a Boolean formula can have either a ``positive'' or ``negative'' truth value).
Using this framework, if the necessary gadgets can be created and arranged in our ABED environment within polynomial time, then ABED is PSPACE-hard.
While it may initially seem unclear as to how exactly this framework can be used to prove PSPACE-hardness, the following sections will describe the function of each gadget, as well as how these gadgets combine together within the framework to apply our described TQBF verification process.

\begin{figure}[t]
  \centering
  \includegraphics[width=0.4\linewidth]{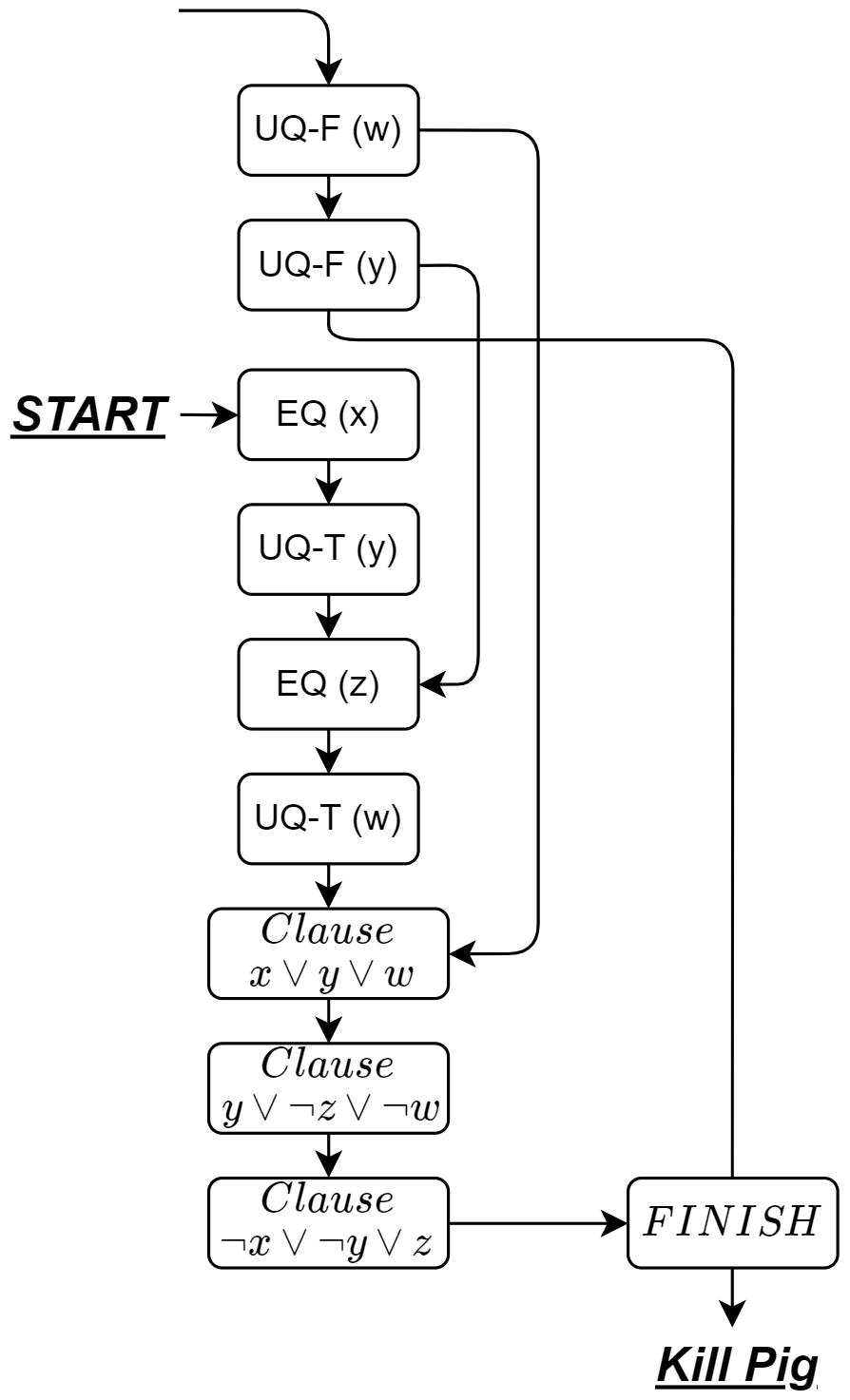}
  \caption{General framework diagram for PSPACE-hardness (ABED).}
\end{figure}

\subsubsection{Formal framework reference terms}
In this section we define some formal terms that can be used to reference specific gadgets within our framework:

\begin{definition}
\emph{(enabled, disabled, current, next, next adjacent, next UQ-F, previous, first, last):}
Each gadget can either be \emph{``enabled''} or \emph{``disabled''} (exactly what this means for each type of gadget is discussed in the next section).
The \emph{``current''} gadget ($Q_{i}$) is the (vertically) lowest enabled gadget in the general framework diagram (Figure 7).
The \emph{``next''} gadget ($Q_{i+1}$) for the current gadget is indicated by the arrows in our general framework diagram, which represent the scope of each quantifier.
For each UQ-F gadget there are two possible next gadgets, the next gadget for the UQ-T gadget associated with its variable (horizontal output arrow in Figure 7) referred to as the \emph{``next adjacent''} gadget, and the UQ-F gadget directly below it (vertical output arrow in Figure 7) referred to simply as the \emph{``next UQ-F''} gadget (note that the last UQ-F gadget has no next UQ-F gadget).
The \emph{``previous''} gadget ($Q_{i-1}$) refers to the most recent current gadget (i.e. essentially the opposite of the next gadget).
We also define the terms \emph{``first''} gadget and \emph{``last''} gadget with respect to the vertical position of specific gadget types in our general framework diagram. The highest of a particular gadget type is the first gadget of that type, whilst the lowest is the last gadget (e.g. for Figure 7, the UQ-F Gadget for the variable $w$ is the first UQ-F gadget, whilst the EQ Gadget for $z$ is the last EQ Gadget). 
\end{definition}

\subsubsection{Gadget design requirements}
In this section we describe the purpose and requirements of the gadgets that will need to be followed by our specific ABED gadget implementations / level construction:

\textbf{EQ gadget:} If an EQ gadget is enabled then the player can use it to set the value of its associated variable to either positive or negative. Doing this disables the EQ gadget and allows the player to enable the next gadget.

\textbf{UQ-T gadget:} If a UQ-T gadget is enabled then it automatically sets the value of its associated variable to positive. The player can then enable the next gadget which also disables the UQ-T gadget.

\textbf{UQ-F gadget:} If a UQ-F gadget is enabled then it alternates between allowing the player to do either of the following two actions: ($A$) the player can set the value of its associated variable to negative, which disables the UQ-F gadget and allows the player to enable the next adjacent gadget; or ($B$) the player can disable the UQ-F gadget and enable the next UQ-F gadget. Note that, as previously mentioned, the last UQ-F gadget does not have a next UQ-F gadget. Attempting to enable the next UQ-F gadget from the last UQ-F gadget will instead attempt to pass through the Finish gadget and solve the level.

\textbf{Clause gadget:} A Clause gadget is ``activated'' if and only if its associated clause is satisfied (i.e. at least one of the literals in the associated clause is true).
The level can be solved if and only if all Clause gadgets can be activated for each possible value combination of all universally quantified variables (abbreviated to UQVC). This means that the level can be solved if and only if the given quantified Boolean formula is true. If the current gadget is a Clause gadget that is both enabled and activated, then the next gadget can be enabled.

\textbf{Finish gadget:} The Finish gadget can be enabled if and only if all Clause gadgets are both enabled and activated.

\subsubsection{Framework design requirements}
The gadget associated with the quantifier with the largest scope (leftmost quantifier in Boolean Formula) is initially enabled (gadget pointed to by Start label in our general framework diagram), with the UQ-T version of the gadget being enabled if it is a universal quantifier, whilst all other gadgets are disabled. The player can enable the first UQ-F gadget at any time, but doing so when the Finish gadget is disabled will put the level into an unsolvable state (prevents the player from ever being able to pass through the Finish gadget). Enabling the first UQ-F gadget also disables all Clause and Finish gadgets.

Essentially, the Finish gadget is used to maintain the ordering of the framework, by automatically making the level unsolvable if the player attempts to open the first UQ-F gadget at any time except after checking that all Clause gadgets are activated (i.e. once we reach the bottom of the framework we start again from the top). 
This action of enabling the first UQ-F gadget begins a new ``framework cycle'', with each framework cycle testing a specific UQVC. 
Once all possible UQVCs have been tested, and assuming that the Finish gadget has not made the level unsolvable, then the player can pass through the Finish gadget and solve the level.

\subsubsection{Framework process summary}
In summary, the player will initially enable and then disable all EQ and UQ-T gadgets, either choosing the value of the associated variable or having it automatically set to positive whilst doing so. The first Clause gadget is then enabled and if it is activated, then the next Clause gadget can also be enabled. If all Clause gadgets are activated then eventually they will all be sequentially enabled, after which the Finish gadget can be enabled as well. The player can then enable the first UQ-F gadget (begin new framework cycle) without putting the level into an unsolvable state, which also disables all Clause and Finish gadgets. Each time a UQ-F gadget is enabled the outcome will alternate between setting the value of the associated variable to negative and then enabling the next adjacent gadget, or enabling the next UQ-F gadget (both outcomes also disable the current UQ-F gadget). This is equivalent to the next adjacent gadget being enabled if the associated variable was positive and the next UQ-F gadget being enabled if the associated variable was negative. If the next adjacent gadget was enabled, then the player can change the values of any variables associated with EQ gadgets after this point in the framework as well as any subsequent UQ-T gadgets setting the value of their associated variable to positive, after which if all Clause gadgets are still activated then the Finish gadget will be enabled again. This process repeats $2^{U}$ times, where $U$ is the number of universal quantifiers in the Boolean formula. Once the player can enable the next UQ-F gadget for all UQ-F gadgets within a single framework cycle (i.e. once all universally quantified variables are negative) a bird will attempt to pass through the Finish gadget. If the player has ensured that they only enabled the first UQ-F gadget when the Finish gadget was enabled, then the bird will successfully pass through the Finish gadget and kill a single pig to solve the level. While this process may initially seem somewhat confusing, following through our framework using this system will confirm that all UQVCs within the quantified Boolean formula are indeed tested.

This means that solving the level is equivalent to finding a solution to the given quantified Boolean formula. Thus, we can show that ABED is PSPACE-hard if the required gadgets can be successfully implemented within the game's environment and the reduction from quantified Boolean Formula to level description can be achieved in polynomial time.

\subsection{Gadget Design}
This section deals with the implementation and arrangement of the necessary framework gadgets for the ABED game environment. 

All Selector and AUT gates within our gadgets are initially closed except for those in the gadget associated with the leftmost quantifier from the Boolean Formula (pointed to by Start label), which will initially have certain gates open corresponding to the gadget's own definition of being enabled, and the Finish gadget which will be discussed later.

\subsubsection{Existential Quantifier (EQ) Gadget}
The structure of the EQ gadget implementation for ABED is shown in Figure 8. This gadget is comprised of two Selector gates $(S_{1},S_{2})$ and four AUT gates $(A_{1},A_{2},A_{3},A_{4})$, where all AUT gates have traverse paths that can be shot into by the player. An EQ gadget is enabled if $A_{1}$, $A_{2}$, $S_{1}$ and $S_{2}$ are open, otherwise it is disabled. A truth table for this gadget is shown in the Appendix (Figure C.32).

\begin{figure}
  \centering
  \includegraphics[width=0.7\linewidth]{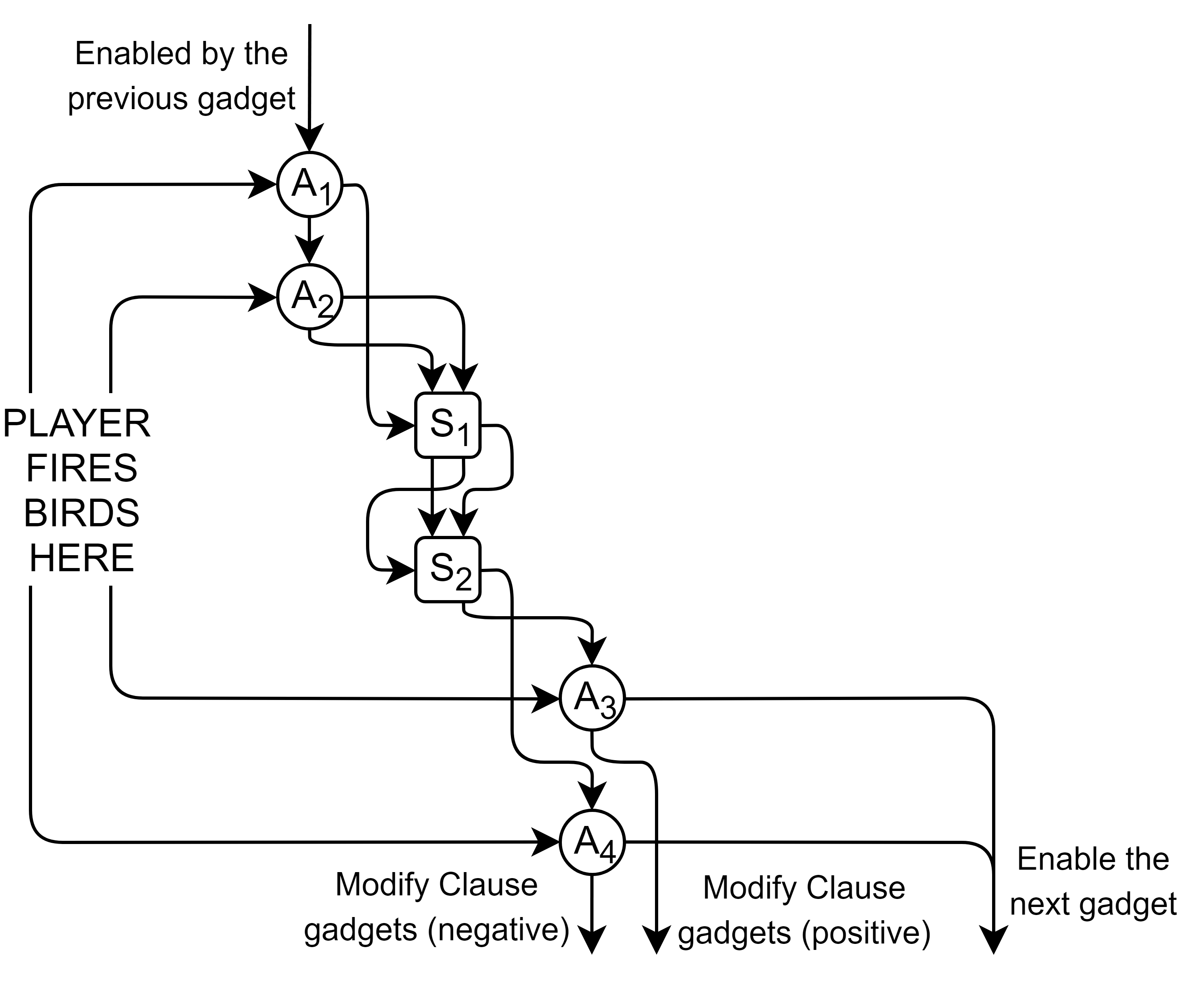}
  \caption{Structure of the Existential Quantifier (EQ) gadget.}
\end{figure}

\begin{lemma}
An EQ gadget can be used to select one of two binary choices, positive or negative, for an associated variable, if and only if it is enabled.
\end{lemma}

\begin{hproof}
AUT gates $A_{1}$ and $A_{2}$ are used to indicate the choice of which value to set the associated variable to. The player fires a bird into the traverse path of $A_{1}$ to indicate a positive value, and $A_{2}$ to indicate a negative value. Traversing $A_{1}$ results in $A_{1}$ and $S_{2}$ being closed and $A_{3}$ being opened, while traversing $A_{2}$ results in $A_{2}$ and $S_{1}$ being closed and $A_{4}$ being opened. Opening either $A_{3}$ or $A_{4}$ sets the value of the associated variable to either positive or negative respectively. 

As the traverse path of $A_{2}$ directly leads into the close path of $S_{1}$, and the traverse path of $A_{1}$ leads into the close path of $S_{2}$ (albeit through $S_{1}$ first), it is impossible to have $A_{2}$ open and $S_{1}$ closed, $S_{1}$ open and $A_{2}$ closed, or $A_{1}$ open and $S_{2}$ closed. The value of the associated variable can only be set to positive by opening $A_{3}$. This can only be done by traversing $A_{1}$ if both it and $S_{1}$ are open. Likewise, the value can only be set to negative by opening $A_{4}$, which is only possible if both $A_{2}$ and $S_{2}$ are open. 

Thus, by combining all this information we can see that neither $A_{3}$ nor $A_{4}$ can be opened if the gadget is disabled. Therefore, the player can only choose the value of the associated variable if the EQ gadget is enabled.
\end{hproof}

\begin{lemma}
An EQ gadget will become disabled after selecting a value for the associated variable.
\end{lemma}

\begin{hproof}
As $A_{1}$ and $A_{2}$ are AUT gates, we know that traversing either of them will close the gate, and thus disable the EQ gadget. Traversing either of these two gates is the only way of selecting a value for the associated variable, so the EQ gadget will clearly be disabled after doing so.
\end{hproof}

\begin{lemma}
The next gadget after an EQ gadget can be enabled if and only if a value has been selected for the associated variable.
\end{lemma}

\begin{hproof}
The next gadget is enabled by firing a bird into the traverse path of either $A_{3}$ or $A_{4}$. Opening either $A_{3}$ or $A_{4}$ sets the value of the associated variable to either positive or negative respectively. Therefore, the value for the associated variable must be selected before the next gadget can be enabled.
\end{hproof}

Essentially, traversing gate $A_{1}$ or $A_{2}$ is used to set the value for the associated variable to either positive or negative respectively (i.e. setter gates). Traversing gate $A_{3}$ or $A_{4}$ is used to enable the next gadget once the player has chosen the value of the associated variable (i.e. checker gates). Which of these two gates ($A_{3}$ or $A_{4}$) is used to achieve this is based on which value was selected for the associated variable, and traversing either gate achieves the same end result. Gates $S_{1}$ and $S_{2}$ ensure that the player can only indicate a single value for the associated variable each time the EQ gadget is enabled. 

To summarise, for each existential quantifier in the given quantified Boolean formula there will be an associated EQ gadget. If an EQ gadget is enabled then the player can use it to set the value of its associated variable to either positive or negative, after which the EQ gadget is disabled and the next gadget is enabled. Once the value of a variable associated with an EQ gadget has been set, it cannot be changed during this framework cycle. The only time the value of an existentially quantified variable can be changed (i.e. its associated EQ gadget is re-enabled), is if it is within the scope of a universal quantifier that has its value changed (perhaps not immediately but will occur before the clauses are next checked for activation).

\subsubsection{Universal Quantifier True (UQ-T) Gadget}
The structure of the UQ-T gadget implementation for ABED is shown in Figure 9. This gadget is comprised of a single AUT gate $(A_{1})$, that has a traverse path which can be shot into by the player. A UQ-T gadget is enabled if $A_{1}$ is open, otherwise it is disabled. A truth table for this gadget is shown in the Appendix (Figure C.33).

\begin{figure}
  \centering
  \includegraphics[width=0.55\linewidth]{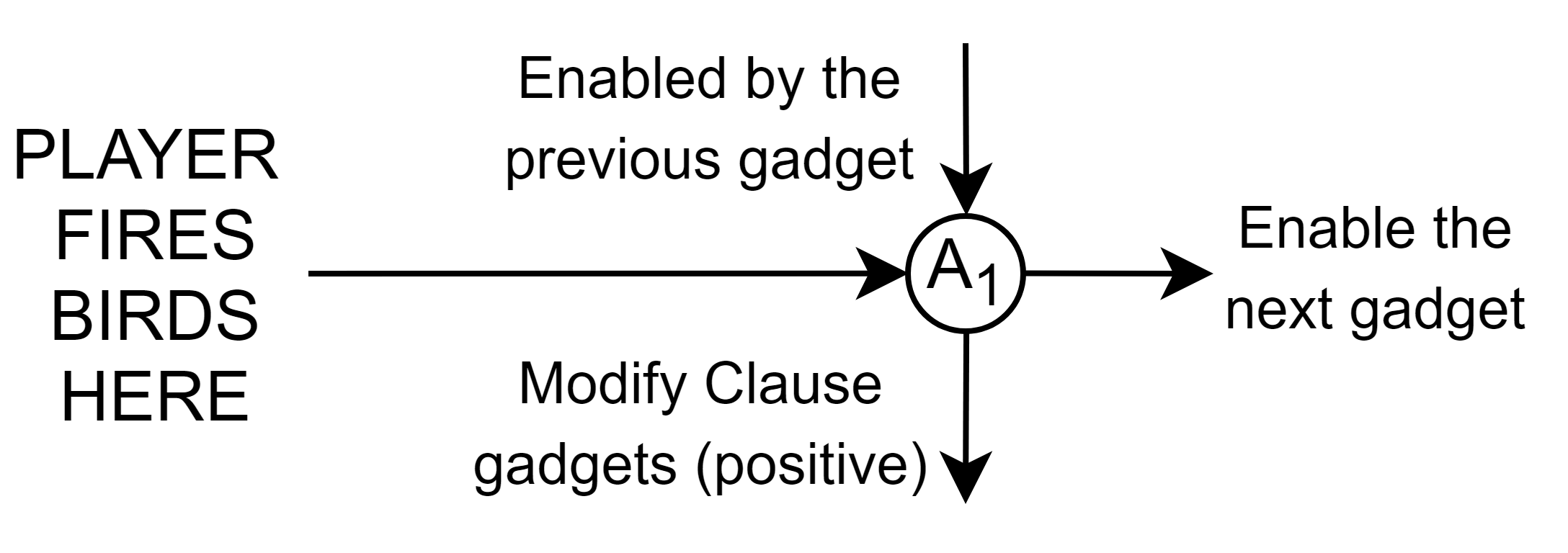}
  \caption{Structure of the Universal Quantifier True (UQ-T) gadget.}
\end{figure}

\begin{lemma}
A UQ-T gadget will set the value of the associated variable to positive, if and only if it is enabled.
\end{lemma}

\begin{hproof}
Opening $A_{1}$ is the only way to enable the gadget, and doing so automatically sets the value of the associated variable to positive.
\end{hproof}

\begin{lemma}
A UQ-T gadget will become disabled after the associated variable has been set to positive.
\end{lemma}

\begin{hproof}
Although the value for the associated variable is automatically set to positive when the gadget is enabled, the player cannot enable any more gadgets until they traverse $A_{1}$. Doing this closes $A_{1}$ and thus disables the gadget. 
\end{hproof}

\begin{lemma}
The next gadget after a UQ-T gadget can be enabled if and only if the associated variable has been set to positive.
\end{lemma}

\begin{hproof}
The next gadget is enabled by firing a bird into the traverse path of $A_{1}$. As opening $A_{1}$ sets the value of the associated variable to positive, this must clearly have already been done in order for the player to traverse $A_{1}$. 
\end{hproof}

\subsubsection{Universal Quantifier False (UQ-F) Gadget}
The structure of the UQ-F gadget implementation for ABED is shown in Figure 10. This gadget is comprised of two Selector gates $(S_{1},S_{2})$ and three AUT gates $(A_{1},A_{2},A_{3})$, where $A_{1}$, $S_{1}$ and $A_{3}$ have traverse paths that be shot into by the player. A UQ-F gadget is enabled if $A_{1}$, $S_{1}$ and $S_{2}$ are open, otherwise it is disabled. A UQ-F gadget is ``unlocked'' if $A_{2}$ is open, otherwise it is ``locked''. Enabling the first UQ-F gadget also disables all Clause and Finish gadgets. A truth table for this gadget is shown in the Appendix (Figure C.34).

 \begin{figure}
  \centering
  \includegraphics[width=0.7\linewidth]{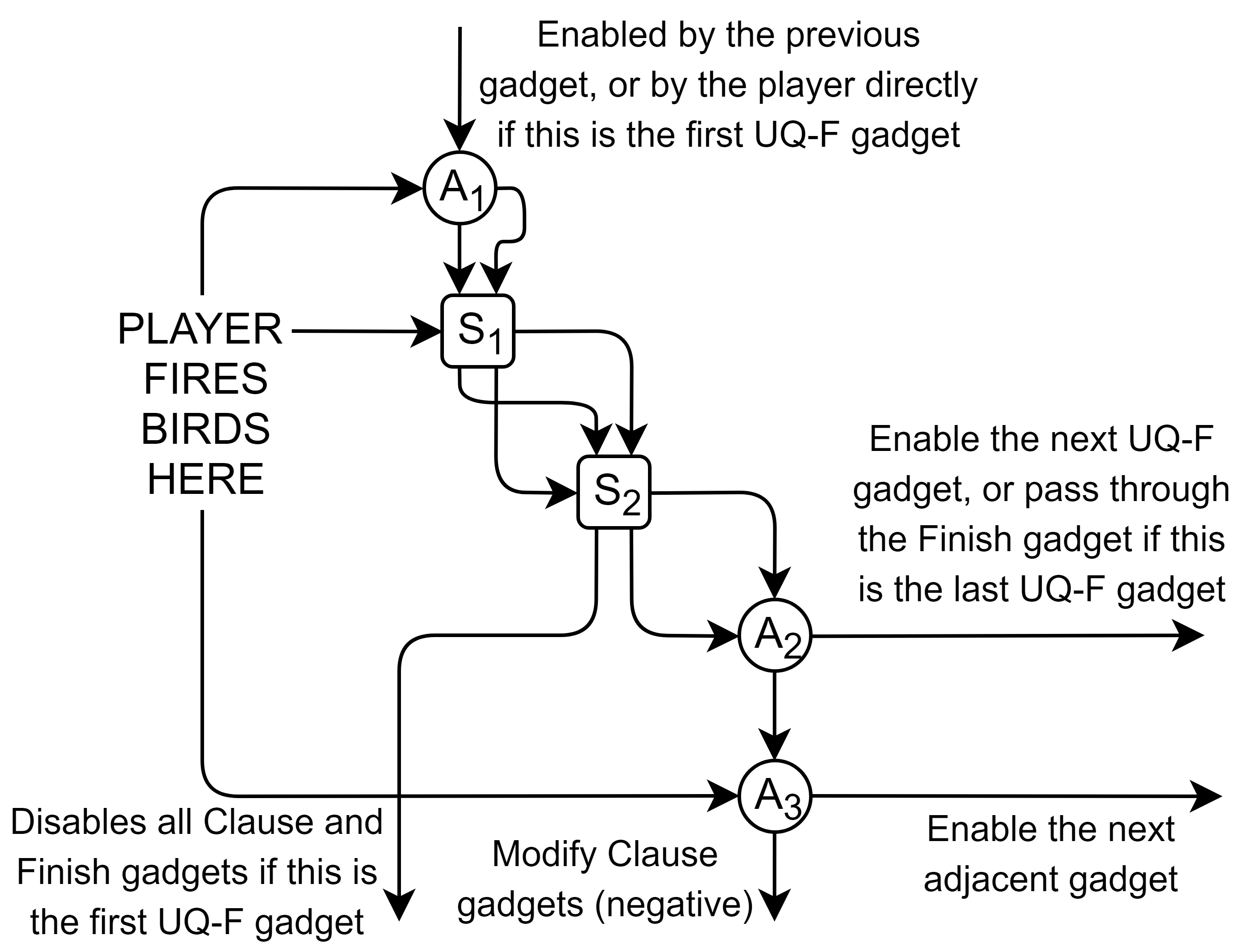}
  \caption{Structure of the Universal Quantifier False (UQ-F) gadget.}
\end{figure}

\begin{lemma}
A UQ-F gadget can be used to set the value of an associated variable to negative, if and only if it is enabled.
\end{lemma}

\begin{hproof}
The only initial thing that a player can do to with a UQ-F gadget after it has been enabled is to traverse either $A_{1}$ or $S_{1}$. Traversing $S_{1}$ would be pointless at this stage as $A_{2}$ is not yet open, so all that would happen is that $S_{2}$ would be closed. Traversing $A_{1}$ instead would close both $A_{1}$ and $S_{1}$ but would also open $A_{2}$ and $A_{3}$, as well as setting the value of the associated variable to negative. 

As the traverse path of $A_{1}$ directly leads into the close path of $S_{1}$ it is impossible to have one open/closed and not the other (both gates must always be in the same position).  If both are closed then the player cannot open $A_{2}$ and $A_{3}$. If $S_{2}$ is closed then it cannot be traversed which also means the player cannot open $A_{2}$ or $A_{3}$. Thus, the value of the associated variable can only be set to negative if the gadget is enabled.
\end{hproof}

\begin{lemma}
A UQ-F gadget will become disabled and unlocked after the associated variable has been set to negative.
\end{lemma}

\begin{hproof}
The only way to set the value of the associated variable to negative is to open $A_{3}$. The only way to achieve this is to traverse $A_{1}$, which closes both $A_{1}$ and $S_{1}$ as well as opening $A_{2}$, causing the UQ-F gadget to be both disabled and unlocked.
\end{hproof}

\begin{lemma}
The next adjacent gadget after a UQ-F gadget can be enabled if and only if the associated variable has been set to negative.
\end{lemma}

\begin{hproof}
Traversing $A_{3}$ is the only way to enable the next adjacent gadget. As opening $A_{3}$ sets the value of the associated variable to negative, this must clearly have already been done first in order for the player to traverse $A_{3}$. 
\end{hproof}

\begin{lemma}
The next UQ-F gadget after a UQ-F gadget can be enabled if and only if the (current) UQ-F gadget is both enabled and unlocked.
\end{lemma}

\begin{hproof}
The only way to enable the next UQ-F gadget is to traverse $A_{2}$ via $S_{1}$. After the player has just unlocked a UQ-F gadget they cannot traverse $A_{2}$ as $S_{1}$ has been closed. Instead they must go back through the framework again, starting from the next adjacent gadget, which can be enabled by traversing $A_{3}$. Once the UQ-F gadget is enabled again the player can then traverse $S_{1}$ (as $A_{2}$ is now open) which enables the next UQ-F gadget (or attempts to pass through the Finish gadget). Traversing $A_{1}$ instead would just result in the same outcome as the first time the gadget was enabled and so would be a redundant action.
\end{hproof}

\begin{lemma}
A UQ-F gadget will become disabled and locked after the next UQ-F gadget is enabled.
\end{lemma}

\begin{hproof}
The only way to enable the next UQ-F gadget is to traverse $S_{1}$. Doing so clearly results in $S_{2}$ and $A_{2}$ being closed in the process (disables and locks the gadget). The player cannot re-open $A_{2}$ as $S_{2}$ is now closed, so the gadget will remain locked until it is re-enabled.
\end{hproof}

Essentially, traversing gate $A_{1}$ is used to set the value of the associated variable to negative, while traversing gate $S_{1}$ is used enable the next UQ-F gadget. The specific wiring arrangement of these gates, along with the gate $S_{2}$, ensures that the player can only select one of these two options each time the UQ-F gadget is enabled. Gate $A_{2}$ ensures that the player can only enable the next UQ-F gadget every other time the current UQ-F gadget is enabled. Traversing gate $A_{3}$ is used to enable the next adjacent gadget, if the player has set the value of the associated variable to negative (i.e. traversed $A_{1}$ instead of $S_{1}$).

To summarise, for each universal quantifier in the given quantified Boolean formula, there will be both an associated UQ-T gadget and UQ-F gadget. Each time the UQ-T gadget is enabled there is only one possible outcome: the value of its associated variable is set to positive, the UQ-T gadget is disabled and the next gadget is enabled. Each time the UQ-F gadget is enabled there are two possible outcomes: ($A$) the value of its associated variable is set to negative, the UQ-F gadget is disabled and the next adjacent gadget is enabled, or ($B$) the UQ-F gadget is disabled and the next UQ-F gadget is enabled (or attempt to pass through the Finish gadget if this is the last UQ-F gadget). The player can always choose outcome $A$, but can only choose outcome $B$ if outcome $A$ was chosen the last time the UQ-F gadget was enabled. However, choosing outcome $A$ when outcome $B$ is possible will never yield a better result, and will only lead to repeat checks of already tested UQVCs. Assuming that the player always selects outcome $B$ whenever they can, each UQ-F gadget will alternate between outcomes $A$ and $B$ each time it is enabled.

\begin{figure}
  \centering
  \includegraphics[width=0.75\linewidth]{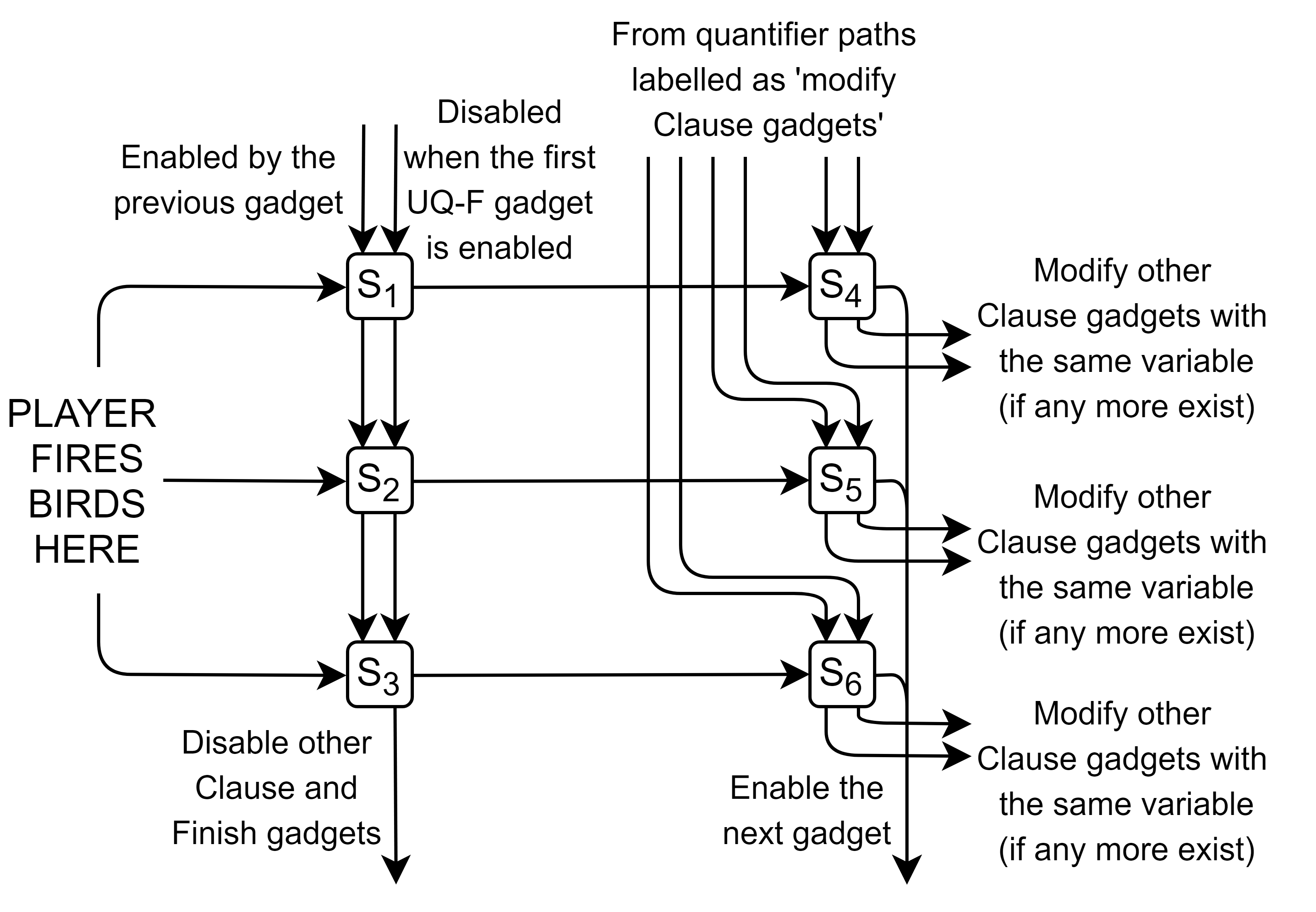}
  \caption{Structure of the Clause gadget.}
\end{figure}

\subsubsection{Clause Gadget}
The structure of the Clause gadget implementation for ABED is shown in Figure 11. This gadget is comprised of six Selector gates $(S_{1},S_{2},S_{3},S_{4},S_{5},S_{6})$, where $S_{1}$, $S_{2}$ and $S_{3}$ have traverse paths that can be shot into by the player. Selector gates $S_{1}$, $S_{2}$ and $S_{3}$ must always be in the same position (closed or open). A Clause gadget is enabled if $S_{1}$, $S_{2}$ and $S_{3}$ are all open, and is disabled if $S_{1}$, $S_{2}$ and $S_{3}$ are all closed. Each Clause gadget is associated with a particular clause from the quantified Boolean formula, and each of the Selector gates $S_{4}$, $S_{5}$ and $S_{6}$ is associated with a specific literal from that clause. The first Clause gadget is enabled by the last Quantifier gadget and the Finish gadget is enabled by the last Clause gadget. A truth table for this gadget is shown in the Appendix (Figure C.35).

When the value of a variable is modified using a Quantifier gadget (exit paths labelled as ``modify Clause gadgets''), the bird on this path will fall down tunnels which lead to the first Clause gadget that contains the variable associated with it. If the value of the variable was set to positive then the bird opens any of $S_{4}$, $S_{5}$ or $S_{6}$ that are associated with the variable's positive literal, whilst closing any of those that are associated with the variable's negative literal (vice versa if the value of the variable was set to negative). This bird then travels into the next Clause gadget that contains this variable, and the process repeats until all applicable Clause gadgets have been visited. Therefore each Clause gadget represents a chosen clause from our quantified Boolean formula, and Selector gates $S_{4}$, $S_{5}$ and $S_{6}$ are either open or closed depending on whether their associated literal is true or not. Therefore, we can say that a Clause gadget is activated if and only if any of $S_{4}$, $S_{5}$ or $S_{6}$ are open.

\begin{lemma}
The next gadget after a Clause gadget can be enabled if and only if the Clause gadget is enabled and activated.
\end{lemma}

\begin{hproof}
The next gadget after a Clause gadget is enabled by firing a bird into the traverse path of $S_{1}$, $S_{2}$ or $S_{3}$. This shot will only enable the next gadget if $S_{4}$, $S_{5}$ or $S_{6}$ is open respectively. This means that at least one of $S_{4}$, $S_{5}$ or $S_{6}$ must be open (i.e. the Clause gadget must be activated) in order for the player to enable the next gadget. This obviously cannot be performed if the Clause gadget is disabled. 
\end{hproof}

To summarise, a player can only enable the next Clause gadget (or enable the Finish gadget if this is the last Clause gadget) if at least one of the literals within the current Clause gadget is true, and thus the clause is activated. Enabling the Finish gadget can therefore only be achieved if all Clause gadgets are activated by the current combination of variable values (i.e. all clauses in the quantified Boolean formula are satisfied).

\subsubsection{Finish Gadget}
The structure of the Finish gadget implementation for ABED is shown in Figure 12. This gadget is comprised of a Selector gate $(S_{1})$ and an AUT gate $(A_{1})$, but the player cannot directly fire into either of them. Traversing $S_{1}$ can also be referred to as ``passing through'' the Finish gadget, and results in the level being solved.
The Finish gadget can exist in one of three states: enabled, disabled and unsolvable. 
The Finish gadget is enabled if $A_{1}$ is open and $S_{1}$ is open, disabled if $A_{1}$ is closed and $S_{1}$ is open, and unsolvable if $S_{1}$ is closed. 
The Finish gadget is initially disabled ($A_{1}$ is closed and $S_{1}$ is open). A truth table for this gadget is shown in the Appendix (Figure C.36).

\begin{figure}
  \centering
  \includegraphics[width=0.55\linewidth]{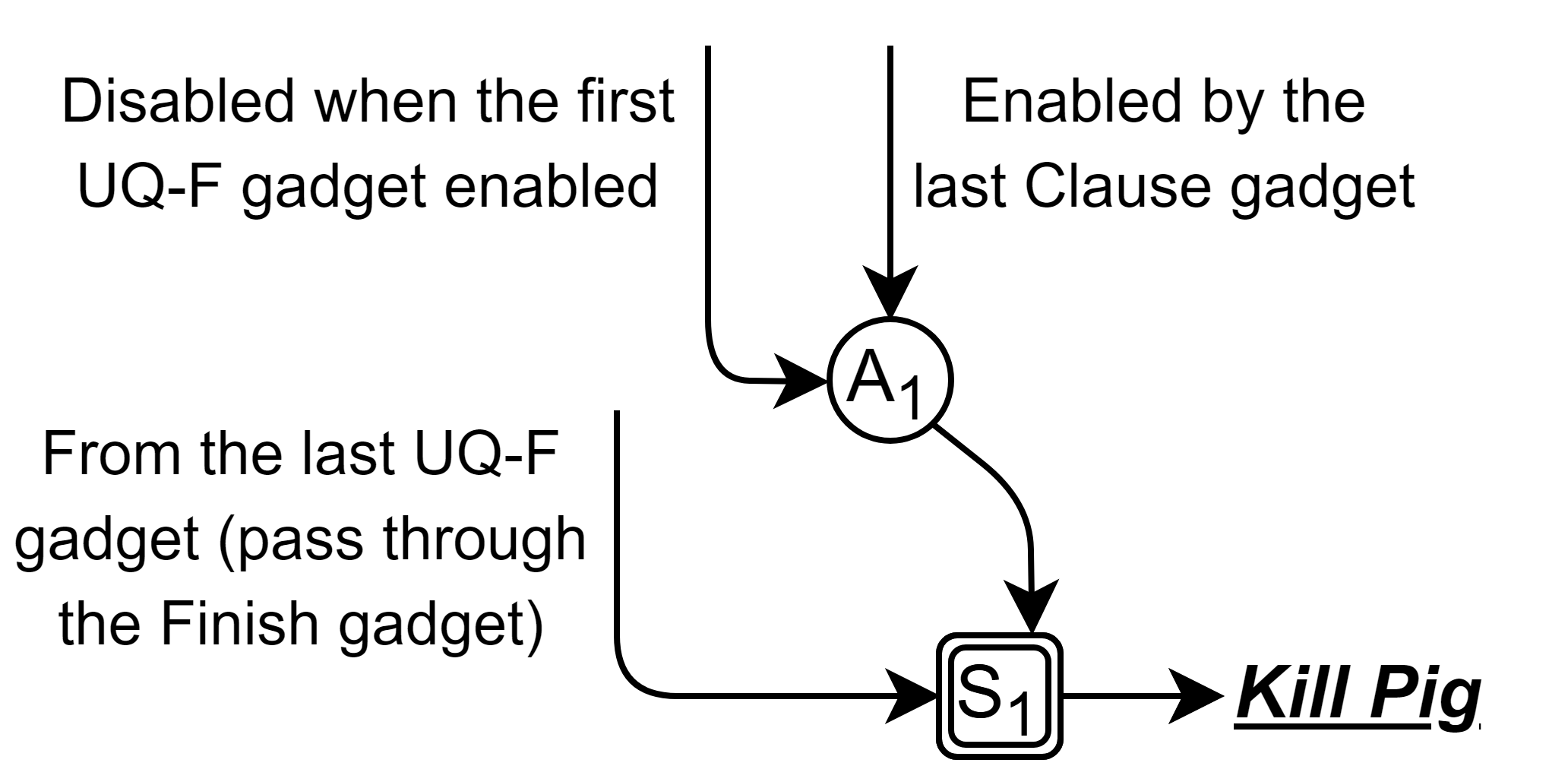}
  \caption{Structure of the Finish gadget.}
\end{figure}

\begin{lemma}
The player can enable the first UQ-F gadget without making the level unsolvable, if and only if the Finish gadget is enabled.
\end{lemma}

\begin{hproof}
The three states that a Finish gadget can be in are all mutually exclusive. Also as there is no way of opening $S_{1}$, if the Finish gadget is ever in the unsolvable state then it can never be taken out of this state. Therefore, as traversing $S_{1}$ is the only way to solve the level, if the Finish gadget is ever unsolvable then the level is unsolvable. While closing $S_{1}$ does not immediately satisfy the loss condition for the level, and allows the player to continue to make further shots, the player can no longer reach the win condition so their loss is guaranteed (eventually the player will run out of birds). We can also observe that the Finish gadget becomes disabled if and only if it is enabled and the first UQ-F gadget is enabled, and that the Finish gadget becomes unsolvable if and only if it is disabled and the first UQ-F gadget is enabled. Therefore, the only way for us to enable the first UQ-F gadget without making the level unsolvable is if the Finish gadget is enabled.
\end{hproof}

Essentially, as the Finish gadget can only be enabled if the last (and by extension all) Clause gadget(s) are enabled and activated, coupled with the fact that opening the first UQ-F gadget disables all Clause and Finish gadgets, we can ensure that the first UQ-F gadget can only be enabled directly after the Clause gadgets have been checked for activation.
Also, as the only way to solve the level to traverse $S_{1}$, which can only happen from the last UQ-F gadget, we can guarantee that all UQVCs are tested before the level can be solved.

\subsection{Level Construction}
This section deals with the reduction process from any given quantified 3-CNF Boolean formula to an equivalent ABED level description, using our previously described framework and gadgets. 
As Angry Birds is a game that relies heavily on physics simulations to resolve player actions, the relative positions of the gadgets is extremely important. Elements within the game are bound by the physics of their environment and the only immediate control the player has is with regard to the shots they make. For this reason, it is necessary to confirm that the gadgets described can be successfully arranged throughout the level space.

\begin{lemmaz}
Any given TQBF problem can be reduced to an ABED level description in polynomial time.
\end{lemmaz}

\begin{zproof}
As each of the necessary gadgets can be created using a constant amount of space and elements, they can also be described in polynomial time. Consequently, the only remaining requirement is that all gadgets can be successfully arranged throughout the level in polynomial time, relative to the size of the quantified 3-CNF Boolean formula. As the number of gadgets required is clearly polynomial, it suffices to describe a polynomial time method for determining the location of each gadget, as well as the level's width, height, slingshot position and number of birds.

Whilst the exact calculations for determining gadget positions for a given quantified Boolean formula can be determined, they are exceptionally long and somewhat irrelevant to this proof. Instead, we will simply show that the tunnels out from each gadget can connect to their appropriate destinations in a polynomial amount of space, and can therefore also be defined in polynomial time. The number of tunnels out of each gadget type is constant, and the number of each gadget type is polynomial. Because of this, there are only a polynomial number of tunnels to consider and each of these can always be connected to their appropriate destination gadget using a polynomial amount of space. This means that the entire framework must also be polynomial in size, and can therefore be described in polynomial time. 
We also know that there are a polynomial number of entrance tunnels to these gadgets that the player can fire into, determined based on the number of quantifier and clause gadgets. Each of these entrance tunnels can simply start above the framework (facing downwards) and then lead into the required gadget entrances. This allows us to define the total width $(W_{T})$ of all entrance tunnels that the player can fire into, which is also polynomial in size.

Although the speed at which a bird can be fired from the slingshot is bounded (less than or equal to a maximum velocity $v_{M}$), we can still ensure that all gadgets are reachable from the slingshot by placing them lower in the level. As there is no air resistance, the trajectory of a fired bird follows a simple parabolic curve for projectile motion, $y=x\tan(\phi)-\frac{g}{2v_{0}^{2}\cos^{2}(\phi)}x^{2}$, where $v_{0}$ is the initial velocity of the fired bird, $\phi$ is the initial angle with which the bird was fired, and $g$ is the gravitational force of the level. While it is highly likely that Angry Birds has a maximum speed that an element could possess, this is not addressed by the formula given (i.e. we assume a theoretical worst case scenario of no terminal velocity). This means that in order for us to ensure that all gadgets are reachable, they must be placed at a distance below the slingshot equal to or greater than $-W_{T}+\frac{g}{v_{M}^{2}}W_{T}^{2}$. We can also use the same formula to calculate the maximum height that a bird fired from the slingshot can reach, $\frac{v_{M}^{2}}{2g}$. Using this we can set the position of the slingshot to $(0,\frac{-v_{M}^{2}}{2g})$ and place all entrance tunnels that the player can fire into the required distance below this in a horizontal alignment against the left side of the level. 
In addition, we need to guarantee that there are enough release points available to allow a bird to be shot into any entrance tunnel for any gadget. To ensure this, we simply move everything constructed so far $W_{T}$ pixels to the right.
Lastly, the number of birds that the player has is equal to $(C+2E+3U)2^U$ (although often this many are not needed), where $C$ is the number of clauses, $E$ is the number of existential quantifiers, and $U$ is the number of universal quantifiers, within the given quantified Boolean formula.
\end{zproof}

An example diagram of a fully constructed structure, using the same quantified Boolean formula as in Figure 7, is shown in the Appendix (Figure A.27).

As we have constructed the necessary gadgets and can position them within the game's environment in polynomial time, the problem of solving levels for ABED is PSPACE-hard.

\begin{theorem}
The problem of solving levels for ABED is PSPACE-hard.
\end{theorem}

\subsection{Winning Strategy (Example)}
We now describe an example of a winning strategy for solving an ABED level description that has been reduced from the same quantified Boolean formula as in Figure 7. For this level description, one strategy that would solve the level would be to set the value of $x$ to positive at the start (after which the EQ gadget associated with $x$ is never enabled again), and set the value of $z$ to be the same as $y$ whenever the EQ gadget associated with $z$ is enabled. The framework will then be cycled four times, for each combination of values for $y$ and $w$, giving the following variable value combinations when the Clause gadgets are enabled:
\begin{itemize}  
\item Framework cycle \#1: $x=1, y=1, z=1, w=1$
\item Framework cycle \#2: $x=1, y=1, z=1, w=0$
\item Framework cycle \#3: $x=1, y=0, z=0, w=1$
\item Framework cycle \#4: $x=1, y=0, z=0, w=0$
\end{itemize}

By comparing these variable values against our quantified Boolean formula, we can see that all clauses are satisfied for each framework cycle, allowing us to enable the Finish gadget and begin the next framework cycle. Essentially, this particular strategy ensures that all Clause gadgets for the given quantified Boolean formula are activated for all UQVCs. As both universally quantified variables ($y$ and $w$) are set to negative on the fourth framework cycle, the fifth framework cycle will allow us to pass through the Finish gadget and solve the level.
A table detailing the 36 shots needed to solve this level is shown in the Appendix (Figure B.31).

\subsection{In PSPACE}
As we have already shown that ABED is PSPACE-hard, the only remaining requirement for completeness is that it also be in PSPACE. The problem of solving levels for ABED can be defined as within PSPACE if it is possible to solve any given level in polynomial space relative to the size of the level's description, and that there are a finite number of states and strategies for solving any given level.

\begin{lemmaz}
Any given ABED level can be solved in polynomial space.
\end{lemmaz}

\begin{zproof}
All game elements can be described using a polynomial amount of memory (e.g. position, velocity, size, etc.), the size of a level does not increase (pre-defined out of bounds limits), no additional elements are added to a level whilst playing (only removed), and every game element behaves deterministically based on a function of the player's actions. Because of this, the current state of a level can always be stored in polynomial space. Thus, the state space of a level can be searched non-deterministically for any possible solutions. This means that the problem is in NPSPACE. We can then use Savitch's theorem \cite{ori10} that NPSPACE = PSPACE to conclude that the problem of solving levels for ABED is indeed in PSPACE.
\end{zproof}

\begin{lemmaz}
There are a finite number of states and strategies for any given ABED level.
\end{lemmaz}

\begin{zproof}
The state of a level is defined based on the current attribute values of all the elements within it. These attribute values are all defined as rational numbers that each take up a finite amount of memory. Therefore, it must also be possible to define the current state of any given level in a finite amount of memory. Thus, the total number of states for any given level is finite. As the number of shots and release points for any given level is polynomial, relative to the size of the level's description, the number of possible strategies for a level is also finite.
\end{zproof}

Thus, as ABED is both PSPACE-hard and in PSPACE, the problem of solving levels for ABED is PSPACE-complete.

\begin{theorem}
The problem of solving levels for ABED is PSPACE-complete.
\end{theorem}

\section{PSAPCE-Hardness of ABPS (polynomial and stochastic)}
\subsection{Framework}
Whilst the problem of solving levels for ABED has been proven PSPACE-complete, it is also possible to show that solving levels for ABPS is PSPACE-hard. This version of Angry Birds no longer allows for an exponential number of birds, but does feature a stochastic game engine. Our proof of PSPACE-hardness for ABPS is based on the same TQBF problem as for ABED, and uses a very similar framework, see Figure 13 (also uses the same example quantified Boolean formula from Figure 7).

\begin{figure}
  \centering
  \includegraphics[width=0.35\linewidth]{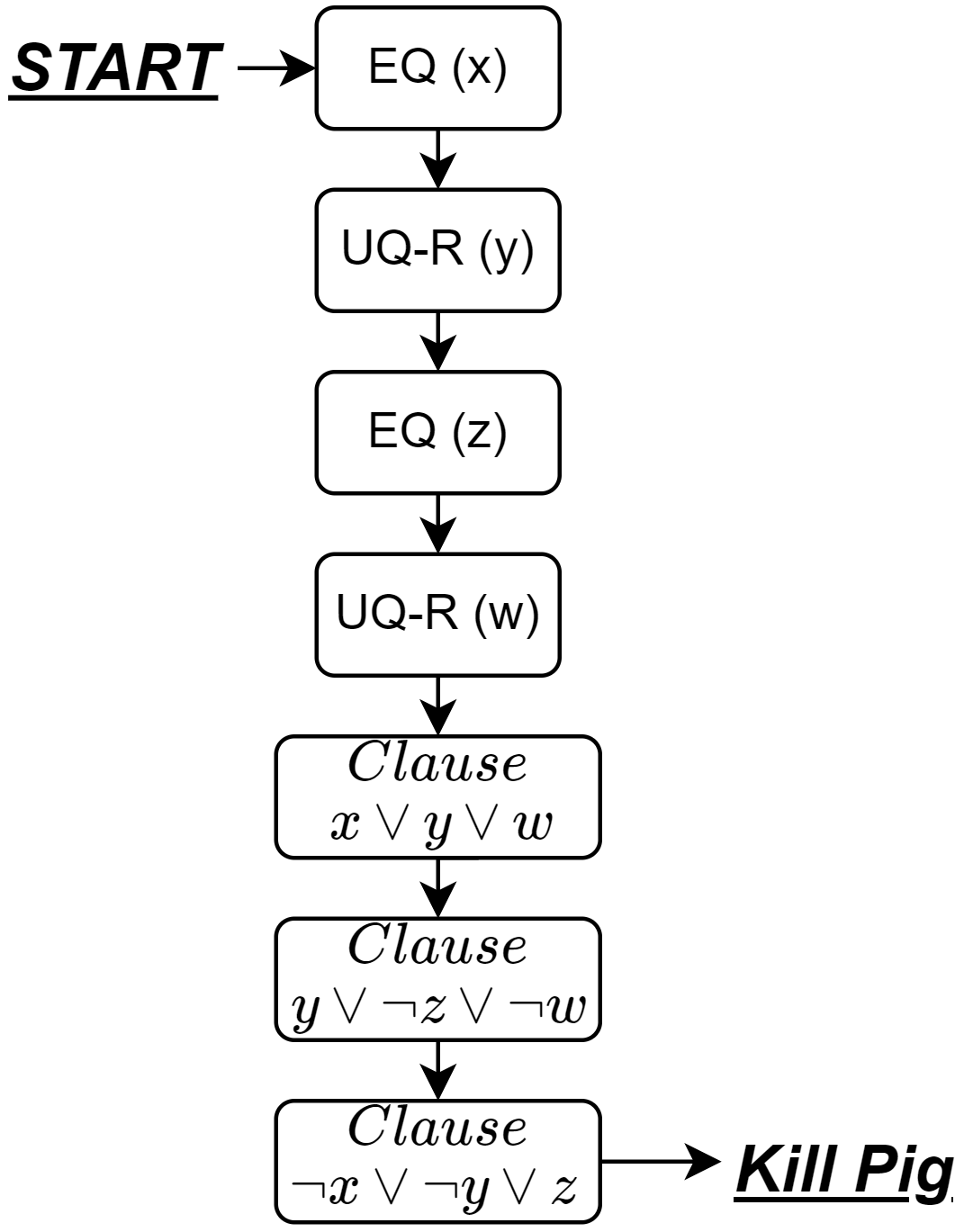}
  \caption{General framework diagram for PSPACE-hardness (ABPS).}
\end{figure}

The EQ and Clause gadgets from the ABED proof remain the same, except that all Clause gadgets are initially set up as if all universally quantified variables are negative. We no longer require UQ-F or Finish gadgets, and UQ-T gadgets are replaced by a new Universal Quantifier Random (UQ-R) gadget. Each UQ-R gadget has a non-zero and non-certain probability of setting the value of its associated variable to positive when it is enabled. If all Clause gadgets are activated after the player has selected a value for each existentially quantified variable, and the value for each universally quantified variable has been (randomly) either set to positive or remains negative, then the player will be able to kill a single pig within the level which replaces the Finish gadget. We also only need as many birds as there are variables and clauses within the given quantified Boolean formula (i.e. the number of birds needed is polynomial).

Essentially, we are no longer testing out every possible UQVC, but are testing a single possible UQVC that is selected at random. As our formal decision problem posed at the beginning of this paper was to determine if there exists a strategy that ALWAYS solves a given level, these two testing approaches are equivalent (as long as the probability of selecting each possible UQVC is greater than zero).

\subsection{Universal Quantifier Random (UQ-R) Gadget}
The structure of the UQ-R gadget implementation for ABPS is shown in Figure 14. This gadget is comprised of an AUT gate $(A_{1})$ and a Random gate $(R_{1})$, where $(A_{1})$ has a traverse path which can be shot into by the player. A UQ-R gadget is enabled if $A_{1}$ is open , otherwise it is disabled. This gadget behaves in a similar manner to the UQ-T gadget from our ABED proof, except that instead of always setting the value of the associated Boolean variable to positive it has a non-zero and non-certain probability of doing so. A truth table for this gadget is shown in the Appendix (Figure C.37).

\begin{lemma}
A UQ-R gadget has a non-zero and non-certain probability of setting the value of an associated variable to positive, if and only if it is enabled.
\end{lemma}

\begin{hproof}
Opening $A_{1}$ is the only way to enable the gadget, and doing this causes a bird to also enter $R_{1}$. This bird then has a non-zero probability of leaving $R_{1}$ through the left exit, but also has a non-zero probability of not leaving $R_{1}$ (either by being trapped in the right exit or by remaining on the point inside the gate). If the bird leaves $R_{1}$ through the left exit then the value of the associated variable is set to positive.
\end{hproof}

\begin{figure}
  \centering
  \includegraphics[width=0.55\linewidth]{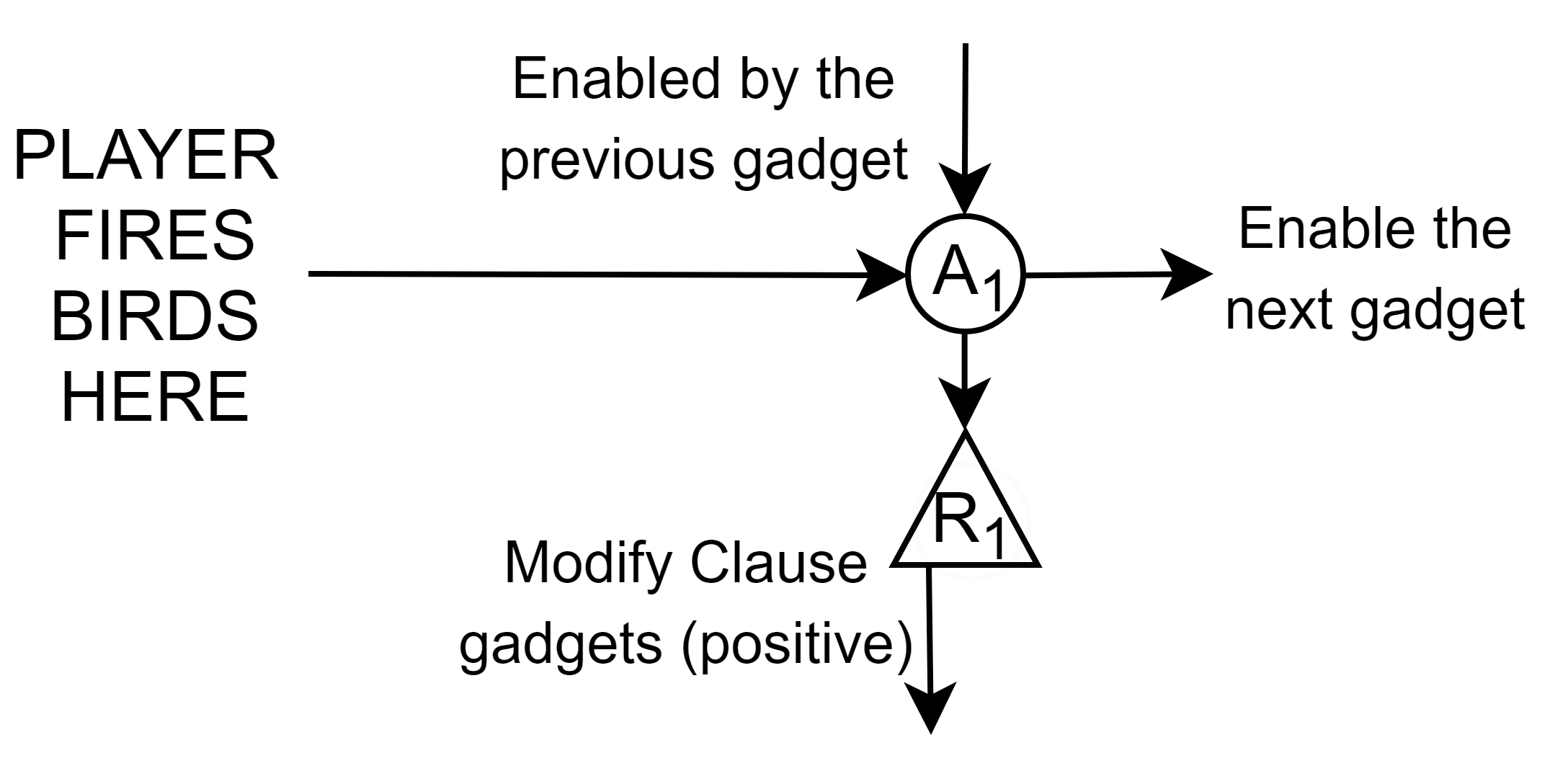}
  \caption{Structure of the Universal Quantifier Random (UQ-R) gadget.}
\end{figure}

Properties and justifications for how the UQ-R gadget is disabled and how the next gadget is enabled can be easily generalised from Section 4.2.2.

Essentially, as all Clause gadgets are initially configured as if all universally quantified variables are negative, when the Clause gadgets are checked for activation there is a non-zero probability that each universally quantified variable will remain negative, but also a non-zero probability that its value will have been changed to positive (i.e. each UQVC has a chance greater than zero of being selected as the outcome).

As the framework for this proof is very similar to that for ABED, the gadgets can be arranged using roughly the same process as described in Section 4.3, except that UQ-T gadgets are replaced by UQ-R gadgets, and no UQ-F or Finish gadgets are necessary. An example diagram of a fully constructed structure, using the same quantified Boolean formula as in Figure 13, is shown in the Appendix (Figure A.28).

As we have constructed the necessary gadgets and can position them within the game's environment in polynomial time, the problem of solving levels for ABPS is PSPACE-hard.

\begin{theorem}
The problem of solving levels for ABPS is PSPACE-hard.
\end{theorem}

\subsection{Winning Strategy (Example)}
The same winning strategy that was used in Section 4.4 ($x=1, z=y$) can also be used here for the same quantified Boolean formula, see Figure 13. In this case, however, the framework does not need to be cycled multiple times to test each UQVC, but instead one of the four possible UQVCs will be randomly selected. As all clauses remain satisfied for our strategy regardless of which UQVC is selected, we can guarantee that the player will always be able to kill the pig and thus solve the level.

\section{NP-Hardness of ABPD (polynomial and deterministic)}
\subsection{Framework}
By using a very similar framework to those used in the last two PSPACE-hard proofs, we can also show that solving levels for ABPD is NP-hard. While this is the ``weakest'' complexity class that is proven in this paper, this version of Angry birds allows for only a polynomial number of birds and features a deterministic physics engine. Our proof of NP-hardness reduces from the NP-complete problem 3-SAT, which involves deciding whether a given 3-CNF Boolean formula is satisfiable. The framework we use for this proof is essentially a reduced version of that used for the TQBF problem, see Figure 15, and is similar to that used for many past platformer games \cite{ori1,ori7,ori2}. Figure 15 uses the Boolean formula $(x \vee y \vee z) \wedge (\neg x \vee y \vee \neg z) \wedge (\neg x \vee \neg y \vee \neg z)$ as an example.

\begin{figure}
  \centering
  \includegraphics[width=0.35\linewidth]{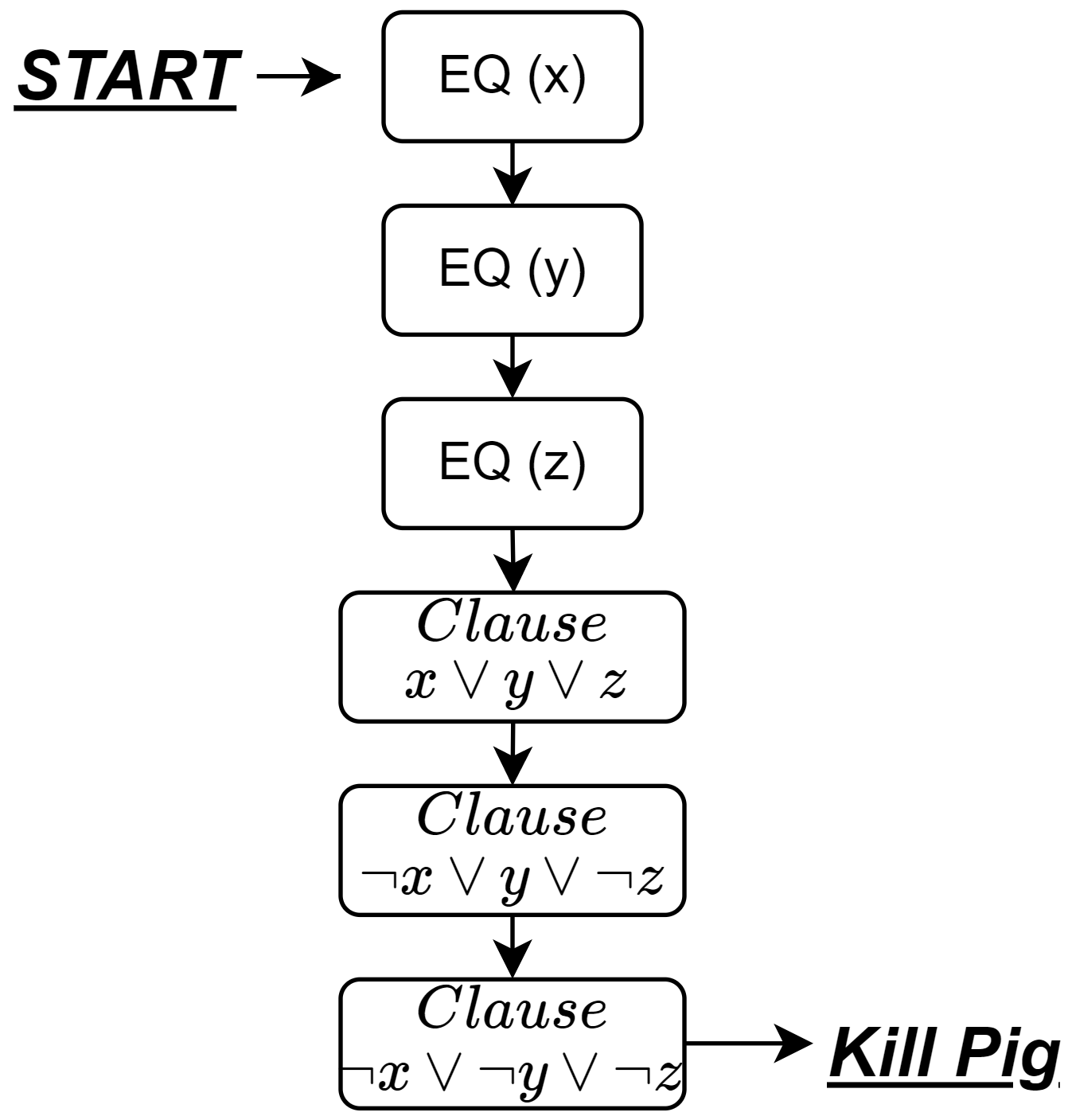}
  \caption{General framework diagram for NP-hardness (ABPD).}
\end{figure}

Essentially any 3-CNF Boolean formula can be represented using our TQBF framework by simply making all variables existentially quantified. This removes the need for any UQ-F, UQ-T or Finish gadgets, relying only on the EQ and Clause gadgets (i.e. for each variable in the Boolean formula there is an associated EQ gadget and for each clause in the Boolean formula there is an associated Clause gadget). If all Clause gadgets are activated after the player has selected a value for each variable, then the player will be able to kill a single pig within the level that replaces the Finish gadget. We also only need as many birds as there are variables and clauses within the given Boolean formula.

As the framework for this proof is very similar to that of ABED, the gadgets can be arranged using roughly the same process as described in Section 4.3, except that no UQ-F, UQ-T or Finish gadgets are necessary. An example diagram of a fully constructed structure, using the same Boolean formula as in Figure 15, is shown in the Appendix (Figure A.29).

As we have constructed the necessary gadgets (although no new gadgets were added for this proof) and can position them within the game's environment in polynomial time, the problem of solving levels for ABPD is NP-hard.

\begin{theorem}
The problem of solving levels for ABPD is NP-hard.
\end{theorem}

We should point out that an NP-hard proof for a version of Angry Birds which had a similar environment to ABPD was previously presented by us in \cite{AIIDE1715829}. However, this proof also used ``breakable blocks'' in addition to the other game elements mentioned in our requirements. This proof was arguably simpler than the one which we present here, but due to the fact that it required additional game elements, we treat this new proof as an improved alternative to that presented in \cite{AIIDE1715829}.

\subsection{Winning Strategy (Example)}
We now describe an example of a winning strategy for solving an ABPD level description that has been reduced from the same quantified Boolean formula as in Figure 15. For this level description, one strategy that would solve the level would be to set the value of $x$ to positive, the value of $y$ to positive, and the value of $z$ to negative. This will ensure that all Clause gadgets are activated, allowing us to kill the pig and solve the level.

\section{EXPTIME-hardness of ABES (exponential and stochastic)}

\subsection{EXPTIME-Complete Original Game}
To show that solving levels for ABES is EXPTIME-hard we will reduce from a known EXPTIME-complete decision problem. For our proof we will use the problem of determining whether a player can force a victory for the game G2, as shown in \cite{exp1}. G2 is a game that is played between two people, with each player attempting to win the game before the other player does. A full and formal definition of G2 can be found in \cite{exp1}, but we provide here a simplified explanation of how it is played.

The game is setup as follows. Each player is given a separate 12-DNF Boolean formula which they are attempting to make true.  Each of the variables that are used in these Boolean formulas are assigned to either player 1 or player 2. The initial values of the variables are also set to either positive or negative. 

The game is played as follows. Each player takes turns making a move (starting with player 1), where they can change the value of at most one variable assigned to them (changing the value of no variables is referred to as ``passing''). The first player to have their Boolean formula ``true'' after making a move wins the game. This victory condition is equivalent to saying that whichever player's Boolean formula is satisfied first wins, but if both players' Boolean formulas are satisfied simultaneously then the player that made the most recent move wins.

If, after the game has been setup, a player can guarantee that they will win regardless of the other player's actions, then that player can force a victory, otherwise they cannot. Determining whether player 1 can force a victory is the known EXPTIME-complete decision problem that we will be using for our proof. 

\begin{problem}
 \problemtitle{\textbf{G2 Formal Decision Problem}}
  \probleminput{12-DNF Boolean formula for each player, variable assignment, initial variable values.}
  \problemquestion{Can player 1 force a victory?}
\end{problem}
From this point on we will refer to player 1 as the ``player'' and player 2 as the ``opponent''.

While many classical two-player games such as Chess, Go and Checkers contain the mechanics necessary to mimic games such as G2, Angry Birds does not on first glance appear to be a suitable choice. Angry Birds is a single-player game and so does not inherently feature an opponent, in the traditional sense, against which to play. However, we can instead use the stochasticity of the physics engine as the opponent we will be facing. This stochasticity allows us to create situations where the player is uncertain about the exact outcome of shots that they make. By utilising this element of uncertainty in shot outcomes, we can create a ``random'' opponent, that will make random moves after each of the player's moves. Even though an opponent that just makes random moves may seem very easy to beat, the complexity of determining whether the player can force a victory for a given G2 instance is the same when facing both an opponent that plays optimally and one that plays randomly, as it is always possible that the random opponent will, by pure chance, actually play optimally (i.e. the player must assume Murphy's Law). Even if the player can beat a random opponent many times for a particular G2 instance, if there exists some small probability that the player will not win then they cannot force a victory (i.e. guaranteeing victory against an opponent that makes random moves is the same as against an opponent that plays perfectly).
Exactly how this simulation of a random opponent by our stochastic physics engine is achieved will be discussed in greater detail later. All that needs to be understood now is that the decision problem we are considering involves determining whether the player can force a victory (i.e. guarantee that they can always solve the level) without knowing exactly how the game's physics will respond to their actions.

\subsection{Framework}
For our proof of EXPTIME-hardness we describe a method of combining several new types of gadget to create an ABES representation for any given setup of the game G2. A framework diagram showing how these gadgets connect within the level space is shown in Figure 16, which uses the example Boolean formulas $(x \wedge \neg y \wedge z) \vee (\neg x \wedge y \wedge w)$ for the player and $(x \wedge y \wedge \neg z) \vee (\neg x \wedge  y \wedge \neg w)$ for the opponent. For each Clause in either the player's or opponent's Boolean formula there is an associated Clause gadget. The framework also contains an Ordering, Random, Choice and Result gadget, the purpose of which will be discussed later.

\begin{figure}[t]
  \centering
  \includegraphics[width=0.5\linewidth]{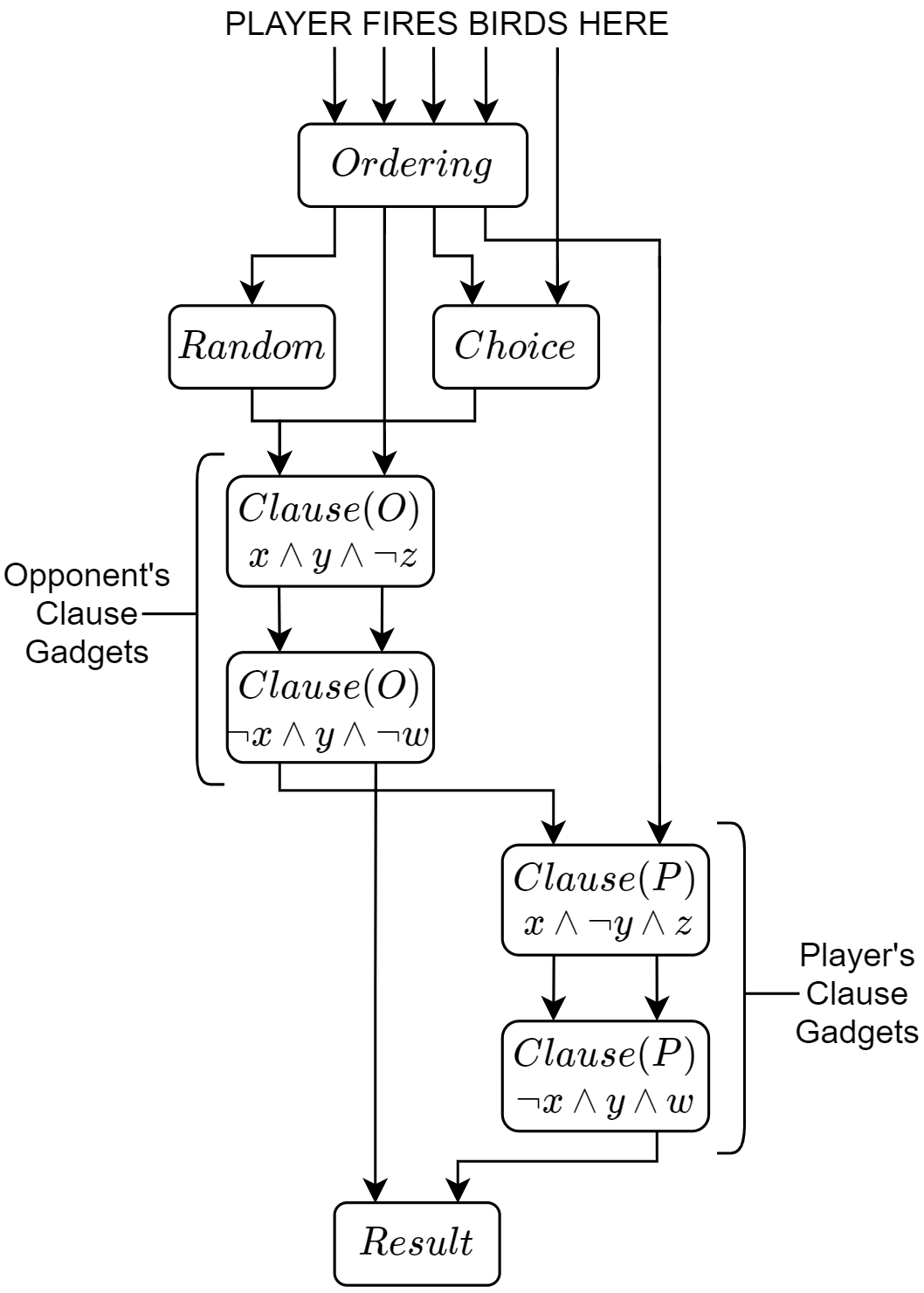}
  \caption{General framework diagram for EXPTIME-hardness.}
\end{figure}

As there is no traditional opponent to make moves for themselves, we must design the level such that the player is forced to make a move for the opponent after they have made their own move. The player first makes their move by either changing the value of a variable assigned to them or by passing. The player can then check whether their own Boolean formula is satisfied, although this is optional and not enforced by the level's design. The player is then forced to randomly change the value of a variable assigned to the opponent (passing is also a possible outcome) and check whether the opponent's Boolean formula is satisfied, before they are allowed to make another move for themselves.

\subsubsection{Gadget design requirements}
The \textit{Ordering} gadget ensures that the correct order of actions is followed by the player. Essentially, all actions must be repeatedly performed in the following order:
\begin{enumerate}  
\item The player makes their move (can effectively skip this step by passing).
\item The player checks whether their Boolean formula is satisfied (can skip this step).
\item The player makes a random move for the opponent (cannot skip but passing may occur as a random possibility).
\item The player checks whether the opponent's Boolean formula is satisfied (cannot skip this step).
\end{enumerate}

The \textit{Choice gadget} allows the player to make a single choice about which of their assigned variables will change in value during their move. The player should also have the option to pass if they do not wish to change the value of a variable. When a bird enters the Choice gadget via the Ordering gadget, the location at which it will exit is based on this choice made by the player. Depending on where the bird exits, the value of a single variable assigned to the player will either be changed or kept the same (pass).

The \textit{Random gadget} makes a random choice between multiple options, based on the stochasticity of the game engine. When a bird enters the Random gadget there are several possible locations where it can exit, each of which has a probability of occurring that is greater than zero. Depending on where the bird exits, the value of a single variable assigned to the opponent will either be changed or kept the same (pass).

Each \textit{Clause gadget} represents a specific clause from either the player's or opponent's Boolean formula, and is ``activated'' if its associated clause is satisfied (i.e. all literals within the associated clause are true). This means that checking if either the player's or opponent's Boolean formula is satisfied, is equivalent to checking if any of their associated Clause gadgets are activated. 
If any of their associated Clause gadgets are activated during this checking step, then a bird will travel into the Result gadget. 
Notions off ``first'', ``last'', ``next'' and ``previous'' Clause gadget are the same as for Section 4.1.

The \textit{Result gadget} is used to decide whether the level has been won or lost, depending on if the player's or opponent's Boolean formula is satisfied first after they have made a move. If the player's Boolean formula is satisfied first, then the player can travel to the Result gadget from one of their activated Clause gadgets, allowing them to ``pass through'' the Result gadget and win the level. If the opponent's Boolean formula is satisfied first, then the player will be forced to travel to the Result gadget from one of the opponent's activated Clause gadgets, which will then close the Result gadget and prevent the player from ever being able to pass through it in the future (i.e. makes the level unsolvable). Essentially, the location and outcome of the first bird to enter the Result gadget depends on whether it came from one of the player's or opponent's Clause gadgets.

\subsubsection{Framework design requirements}
The player fires a bird into the Ordering gadget to make the majority of actions, as well as into the Choice gadget to dictate which of their assigned variables will change in value for their next move. For our general framework diagram (Figure 16), an arrow into the left side of a Clause gadget indicates that the value of a variable is being changed, while an arrow into the right side indicates that the Clause gadget is being checked for activation (i.e. check if associated clause is satisfied). The arrow into the left side of the Result gadget signifies that the level is lost (unsolvable), while the arrow into the right side signifies that the level is won (solved). Lastly, the arrow into the left side of the Choice gadget carries out the player's chosen move, while the arrow into the right side allows the player to specify the move they wish to make next.

This means that solving the level is equivalent to winning a game of G2 (against a random opponent).
Thus, we can show that ABES is EXPTIME-hard if the required gadgets can be successfully implemented within the game's environment and the reduction from G2 setup to level description can be achieved in polynomial time.

\subsection{EXPTIME-Hardness}
This section deals with the implementation and arrangement of the necessary framework gadgets for the ABES game environment, as well as the reduction process from any given setup of G2 to an equivalent ABES level description.

\subsubsection{Ordering Gadget}
The purpose of the Ordering gadget is to ensure that all actions are carried out in the correct order. 
The structure of the Ordering gadget implementation for ABES is shown in Figure 17. This gadget is comprised of two Selector gates $(S_{1},S_{2})$ and an AUT gate $(A_{1})$. $A_{1}$ and $S_{1}$ are initially open while $S_{2}$ is initially closed. There are four entry points to the Ordering gadget $(SO_{I},SP_{I},CO_{I},CP_{I})$ and four corresponding exit points $(SO_{O},SP_{O},CO_{O},CP_{O})$. A bird which enters the Ordering gadget at a given entry point will either leave at the corresponding exit point or fail to leave the Ordering gadget, based on whether certain gates within the Ordering gadget are open or closed. Each exit point leads to the following gadgets/actions: $SP_{O}$ to the Choice gadget (\textbf{P}layer makes their move to \textbf{S}et the truth value for one of their assigned variables), $SO_{O}$ to the Random gadget (make a random move for the \textbf{O}pponent to \textbf{S}et the truth value for one of their assigned variables), $CP_{O}$ to the \textbf{P}layer's \textbf{C}lause gadgets (check whether the player's Boolean formula is satisfied), and $CO_{O}$ to the \textbf{O}pponent's \textbf{C}lause gadgets (check whether the opponent's Boolean formula is satisfied). A deterministic finite state machine (DFSM) showing the relations between gate states, entry points and exit points is shown in Figure 18 (note that the first value given for each arrow is the entry point, while the second value is the exit point; exit points marked as ``-''  indicate that the bird did not leave the Ordering gadget). A truth table for this gadget is shown in the Appendix (Figure C.38).

\begin{figure}
\centering
\begin{minipage}{.36\linewidth}
  \centering
  \includegraphics[width=.99\linewidth]{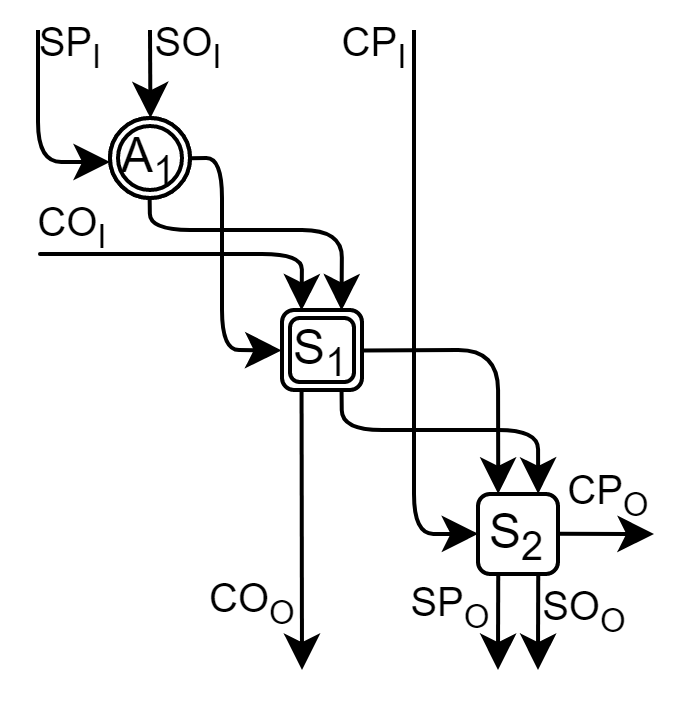}
\captionsetup{justification=centering}
  \captionof{figure}{Model of the Ordering gadget used.}
  \label{fig:test1}
\end{minipage}%
\begin{minipage}{.62\linewidth}
  \centering
  \includegraphics[width=0.99\linewidth]{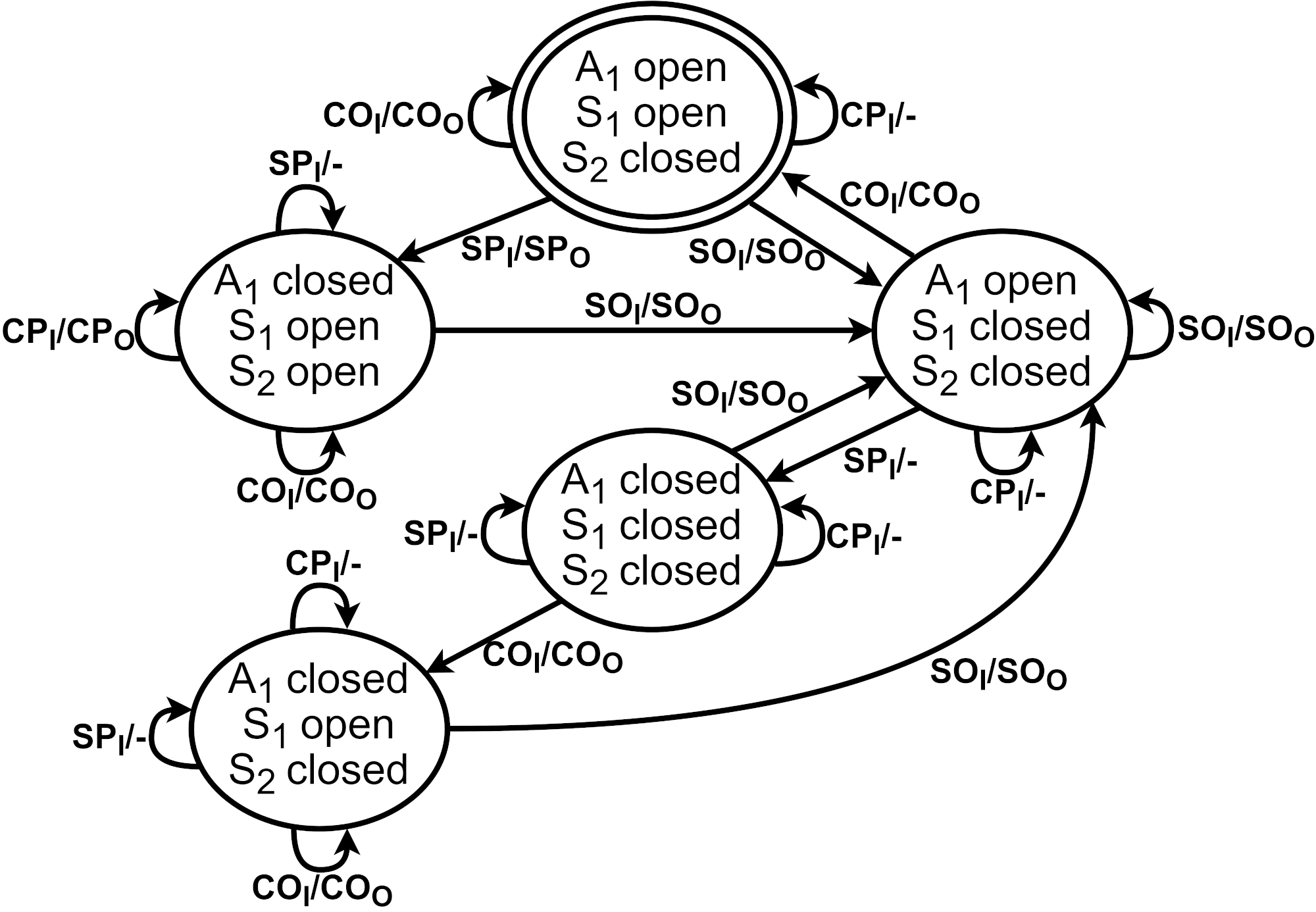}
\captionsetup{justification=centering}
  \captionof{figure}{DFSM for actions performed in Ordering gadget.}
  \label{fig:test2}
\end{minipage}
\end{figure}

Because both the player and opponent can pass as a possible move, and the player does not have to check whether their Boolean formula is satisfied after making their move, we can ensure that the correct order of actions is followed if the following two properties hold.

\begin{lemma}
If the player makes a move, they must make a random move for the opponent and then check whether the opponent's Boolean formula is satisfied, before they can make another move.
\end{lemma}

\begin{hproof}
Using the DFSM in Figure 18, we can see that after a bird exits via $SP_{O}$, a bird must exit via $SO_{O}$ followed by a bird exiting via $CO_{O}$, before a bird can exit via $SP_{O}$ again.
Note that it is also possible for a bird to exit via $SO_{O}$ and/or $CO_{O}$ multiple times before a bird exits via $SP_{O}$ again, but as both the player and opponent have passing as a possible move, there is no issue with this (any duplicate opponent moves can simply be treated as the player passing, and as the opponent can potentially pass their move as a random outcome we only need to check if the opponent's Boolean formula is satisfied if the player didn't pass on their previous move).
\end{hproof}

\begin{lemma}
If the player makes a random move for the opponent, they must check whether the opponent's Boolean formula is satisfied before they can check if the player's Boolean formula is satisfied.
\end{lemma}

\begin{hproof}
Again using the DFSM in Figure 18, we can see that after a bird exits via $SO_{O}$ a bird must also exit via $CO_{O}$, before a bird can exit via $CP_{O}$.
This essentially ensures that the player is only able to check if their Boolean formula is satisfied between making their own move and making a random move for the opponent.
\end{hproof}

\subsubsection{Choice Gadget}
The purpose of the Choice gadget is to allow the player to make a decision about which of their assigned variables will change in value. An example of a Choice gadget implementation for ABES with four possible exit points is shown in Figure 19. This gadget is comprised of a sequence of AUT gates ($A_{1}$, $A_{2}$, $A_{3}$,..., $A_{(2V_{p})}$, where $V_{p}$ is the number of variables assigned to the player). Each AUT gate is associated with a particular value for one of the player's variables (i.e. a literal). The player can directly open any AUT gate within the Choice gadget at any time, and a bird attempts to traverse this sequence of AUT gates whenever it leaves the Ordering gadget from exit $SP_{O}$.

\begin{figure}
  \centering
  \includegraphics[width=0.55\linewidth]{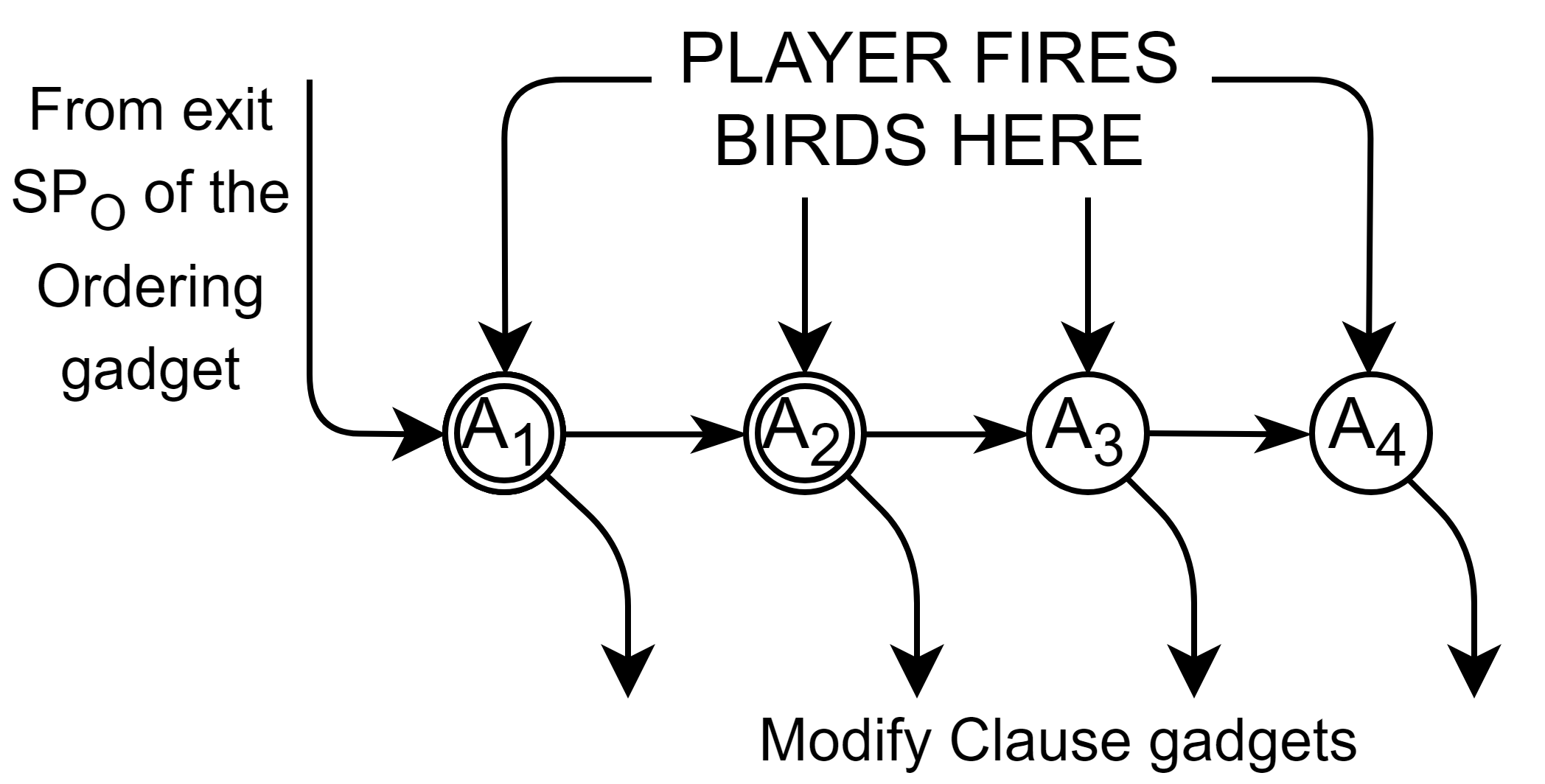}
  \caption{Example model of a Choice gadget with four possible outcomes.}
\end{figure}

\begin{lemma}
The Choice gadget can be used to indicate which of the player's variables will change in value (i.e. which literal to make true).
\end{lemma}

\begin{hproof}
The first AUT gate in the sequence that is closed represents the literal that the player wishes to make true. For the example shown, the player wished to choose the literal represented by the third AUT gate, so has opened all the other AUT gates before it. Essentially, when a bird attempts to traverse this sequence of AUT gates, the first AUT gate that it is unable to traverse represents the selection of its associated literal to make true.
\end{hproof}

\begin{lemma}
A bird which enters the Choice gadget from exit B of the Ordering gadget, will exit the Choice gadget at a location unique to the literal selected by the player.
\end{lemma}

\begin{hproof}
Whilst, the player can open any number of AUT gates within the Choice gadget, they can only be traversed from exit $SP_{O}$ of the Ordering gadget. If an AUT gate is open then a bird can traverse it (closing the AUT gate in the process) and then attempt to traverse the next AUT gate in the sequence. The first AUT gate in this sequence that is closed will prevent the bird from being able to traverse it, meaning it will instead leave the AUT gate at exit $T_{R}$. The bird will then travel into the Clause gadgets and make the desired change, based on the literal associated with this closed AUT gate. The $T_{R}$ exit for each AUT gate in this gadget essentially represents a unique literal that the player can make true during their move, and so the location where a bird exits the gadget is unique to the chosen literal.
\end{hproof}

In summary, the player can determine the exit point for any bird that enters the Choice gadget from exit $SP_{O}$ of the Ordering gadget, by opening all AUT gates before the desired exit point. Each exit point from the Choice gadget then sets the literal associated with its AUT gate to true for both the player's and opponent's Clause gadgets.

\begin{lemma}
The player can pass if they do not wish to change the value of any of their assigned variables.
\end{lemma}

\begin{hproof}
A pass can be made either by selecting a literal that is already true, or by opening all AUT gates in the Choice gadget.
\end{hproof}

\begin{lemma}
The width and height of the Choice gadget, as well as the number of game elements it contains, is polynomial in the number of variables assigned to the player.
\end{lemma}

\begin{hproof}
Let $A_{W}$, $A_{H}$ and $A_{E}$ be constants representing the width, height and number of elements (respectively) for an AUT gate. The width, height and number of elements for a Choice gadget is therefore bounded by the polynomial expressions $(2V_{p})A_{W}$, $(2V_{p})A_{H}$ and $(2V_{p})A_{E}$ respectively.
\end{hproof}

\subsubsection{Random Gadget}
The purpose of the Random gadget is to randomly select one of several options, each of which is associated with a particular value for one of the opponent's variables (i.e. the Random gadget uses the inherent uncertainty in the outcome of collisions to make a random move for the opponent). Each of these options should have a probability greater than zero of occurring, and the player cannot be allowed to influence or know the outcome of the Random gadget in advance. An example of a Random gadget implementation for ABES with four possible exit points is shown in Figure 20. This gadget is comprised of multiple Random gates ($R_{1}$, $R_{2}$, $R_{3}$,..., $R_{(2V_{o}-1)}$), where $V_{o}$ is the number of variables assigned to the opponent, that are arranged in a Binary tree fashion. The first row has one Random gate, then the next two, then four, and so on. A bird enters at the top of this tree of Random gates whenever it leaves the Ordering gadget from exit $SO_{O}$.

\begin{figure}
  \centering
  \includegraphics[width=0.23\linewidth]{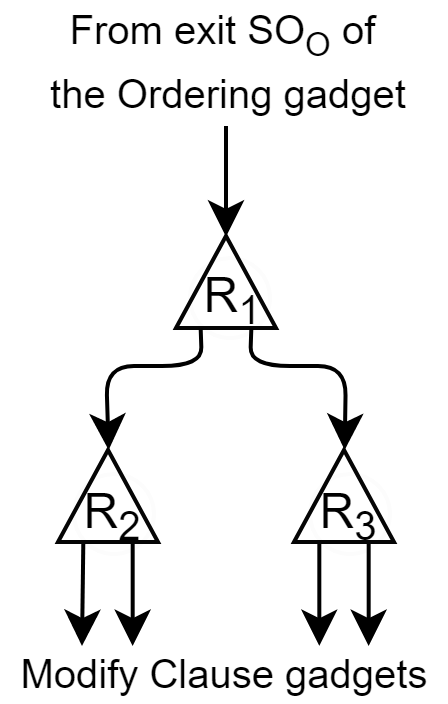}
  \caption{Example model of a Random gadget with four possible outcomes.}
\end{figure}

\begin{lemma}
The Random gadget can be used to randomly select which of the opponent's variables will change in value (i.e. which literal to make true) or pass, using the stochasticity of the game engine.
\end{lemma}

\begin{hproof}
As any bird which enters a Random gate has a probability greater than zero of leaving the Random gate at either exit point, then regardless of how many Random gates the bird interacts with inside our Random gadget, the probability of the bird leaving the Random gadget at any specific exit point is also greater than zero (i.e. by combining together multiple Random gates, it is possible to create a Random gadget that can select between any number of different options). Note that if the bird remains at any point within the Random gadget, then this can simply be treated as a pass.
Each exit point from the Random gadget is associated with a particular literal for one of the opponent's variables. The bird will then travel into the Clause gadgets and make the desired change, based on the literal associated with the exit point. If the literal associated with the bird's exit point is already true then nothing will change (treated as a pass).
\end{hproof}

In summary, any bird that enters the Random gadget from exit $SO_{O}$ of the Ordering gadget has a probability greater than zero of leaving the gadget at any specific exit point. Each exit point from the Random gadget then sets the literal associated with it to true for both the player's and opponent's Clause gadgets.

\begin{lemma}
The width and height of the Random gadget, as well as the number of game elements it contains, is polynomial in the number of variables assigned to the opponent.
\end{lemma}

\begin{hproof}
Let $R_{W}$, $R_{H}$ and $R_{E}$ be constants representing the width, height and number of elements (respectively) for a Random gate. The width, height and number of elements for a Random gadget is therefore bounded by the polynomial expressions $(2V_{o}-1)R_{W}$, $(2V_{o}-1)R_{H}$ and $(2V_{o}-1)R_{E}$ respectively. 
\end{hproof}

\subsubsection{Clause Gadget}
The purpose of the Clause gadget is to represent a single associated clause from either the player's or opponent's Boolean formula, and is activated if the clause is satisfied. An example of a Clause gadget implementation for ABES is shown in Figure 21. This gadget is comprised of a sequence of Selector gates ($S_{1}$, $S_{2}$, $S_{3}$,..., $S_{L}$), where $L$ is the number of literals within its associated clause (maximum of 12). Each of these Selector gates represents a literal from the associated Clause, and is either open or closed depending on whether their associated literal is true or not. Therefore, we can say that a Clause gadget is activated if and only if all Selector gates within it are open. An example truth table for this gadget is shown in the Appendix (Figure C.39).

Figure 22 also provides an example of how multiple Clause gadgets can be combined to represent a complete Boolean formula, in this case for the Boolean formula $(X \wedge Y) \vee (\neg X \wedge \neg Y)$ (i.e. two Clause gadgets which each contain two Selector gates). For this example, the value of $X$ is negative whilst the value of Y is positive. There are five points of entry to the first Clause gadget and the purpose of these different entry points is as follows (starting from the leftmost entry point): check whether any Clause gadgets are activated (if so then bird travels to the Result gadget), set the value of $X$ to positive, set the value of $X$ to negative, set the value of $Y$ to positive, set the value of $Y$ to negative. This arrangement ensures that we can check if any number of Clause gadgets are activated using a single bird.
\begin{figure}
\centering
\begin{minipage}{.52\linewidth}
  \centering
  \includegraphics[width=0.99\linewidth]{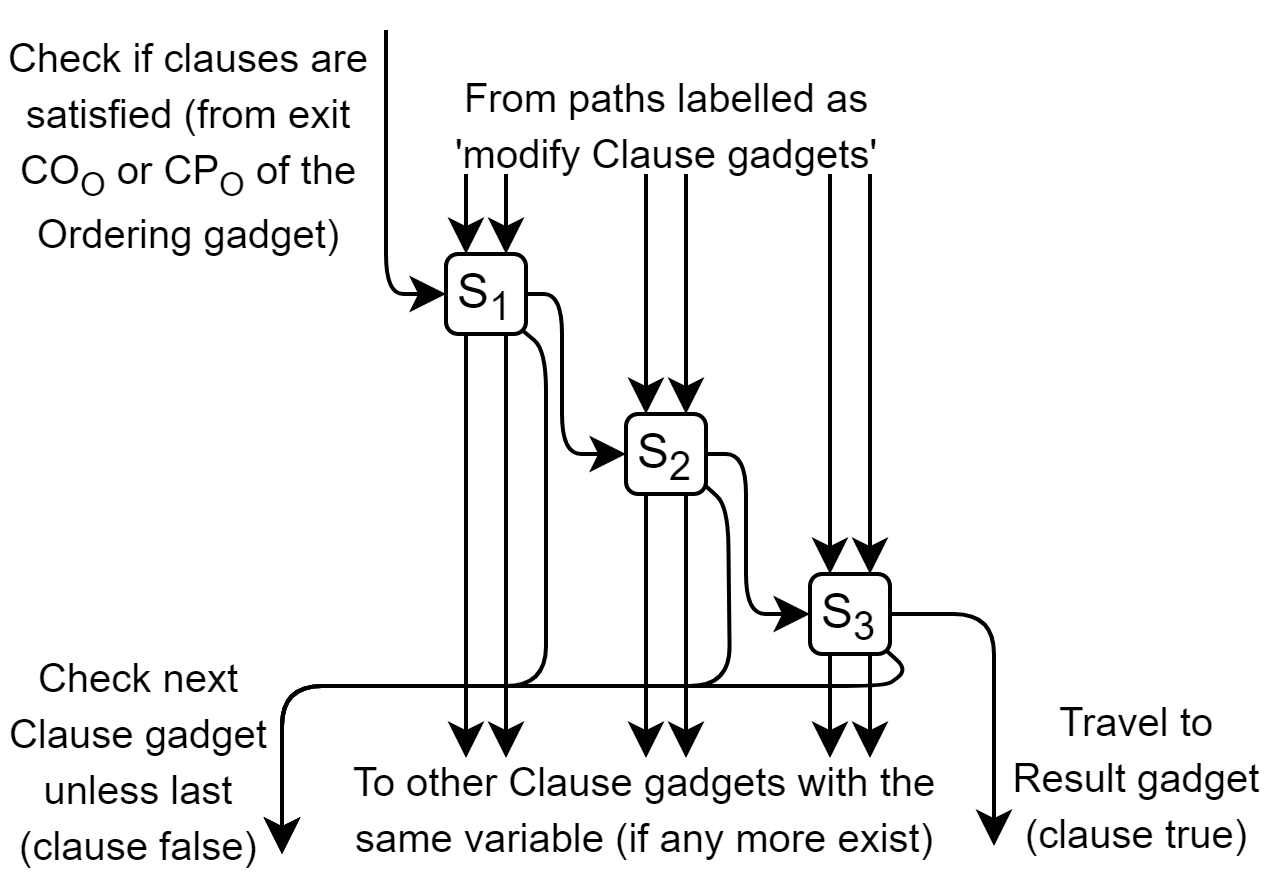}
\captionsetup{justification=centering}
  \captionof{figure}{Example model of a Clause gadget for a Clause with three literals.}
  \label{fig:test2}
\end{minipage}
~
\begin{minipage}{.45\linewidth}
  \centering
  \includegraphics[width=.99\linewidth]{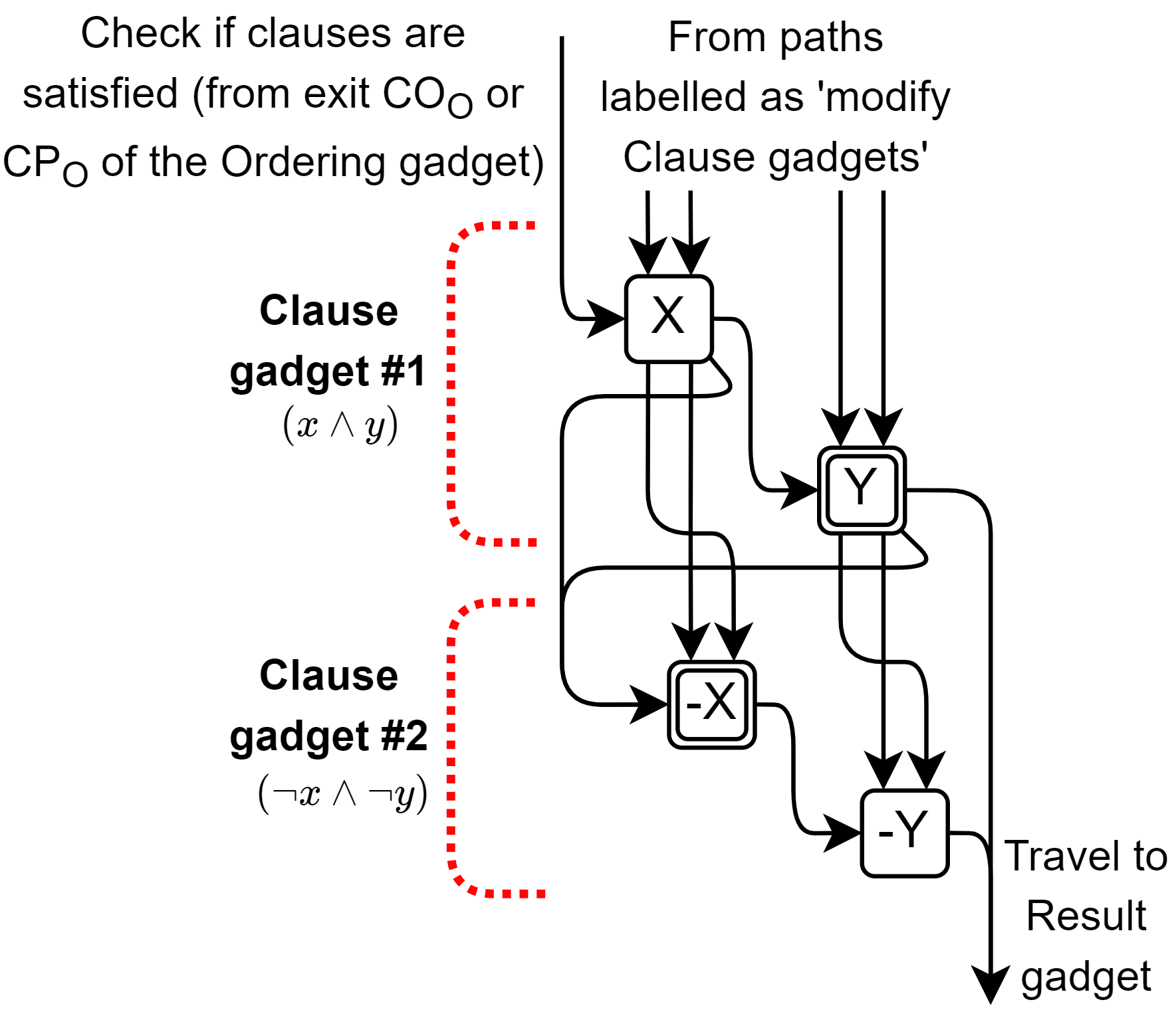}
\captionsetup{justification=centering}
  \captionof{figure}{Example tunnel connection diagram for two Clause gadgets with two literals each.}
  \label{fig:test1}
\end{minipage}
\end{figure}

Whenever the Random or Choice gadget is used to set the value of a variable (exit paths labelled as ``modify Clause gadgets''), a bird will travel through all the Clause gadgets that contain that variable, for both the player and opponent, opening the Selector gates that represent the chosen literal and closing those that represent the negation of it (similar reasoning and setup to the Clause gadget description in Section 4.2.4 for our PSPACE-hard proofs).

\begin{lemma}
The Result gadget can be reached from a specific Clause gadget if and only if the Clause gadget is activated
\end{lemma}

\begin{hproof}
The Result gadget can only be reached from a Clause gadget if a bird is able to traverse every Selector gate within it. As this is clearly only possible if all Selector gates are open, the Clause gadget must be activated for a bird to reach the Result gadget from it.
\end{hproof}

To summarise, each time that we are checking if either the player's or opponent's Boolean formula is satisfied, we are actually sequentially checking if any of the Clause gadgets associated with clauses from their respective Boolean formulas are activated. If any of these Clause gadgets are activated, then a bird will be able to travel to the Result gadget. The location that the bird enters the Result gadget depends on whether the activated Clause gadget that it successfully travelled through was associated with a clause from either the player's or opponent's Boolean formula.

\begin{lemma}
The maximum width and height of a Clause gadget, as well as the number of game elements it contains, is constant.
\end{lemma}

\begin{hproof}
As a 12-DNF Boolean formula can contain a maximum of 12 literals, the maximum number of Selector gates that a Clause gadget can contain is 12. As the width, height and number of elements for each Selector gate is constant, the maximum width, height and number of elements for a Clause gadget is also constant.
\end{hproof}

\subsubsection{Result Gadget}
The purpose of the Result gadget is to either solve the level or make the level unsolvable, depending on whether the player's or opponent's Boolean formula was satisfied first after making their move.
The structure of the Result gadget implementation for ABES is shown in Figure 23. This gadget is comprised of a single Selector gate $(S_{1})$ that is initially in the open position. Traversing $S_{1}$ can also be referred to as passing through the Finish gadget, and results in the level being solved.

\begin{lemma}
The entry point of the first bird to enter the Result gadget will either solve the level or make it unsolvable.
\end{lemma}

\begin{hproof}
If the first bird to enter the Result gadget traverses $S_{1}$, then the bird will kill the pig and solve the level. If the first bird to enter the Result gadget closes $S_{1}$, then the pig can never be killed and the level becomes unsolvable.
\end{hproof}

Because of this, we can simply connect the tunnels so that any bird which enters the Result gadget from one of the player's activated Clause gadgets attempts to traverse $S_{1}$ (i.e. attempts to pass through the Result gadget), and any bird which enters the Result gadget from one of the opponent's activated Clause gadgets closes $S_{1}$ (i.e. makes the level unsolvable).

\begin{figure}
  \centering
  \includegraphics[width=0.5\linewidth]{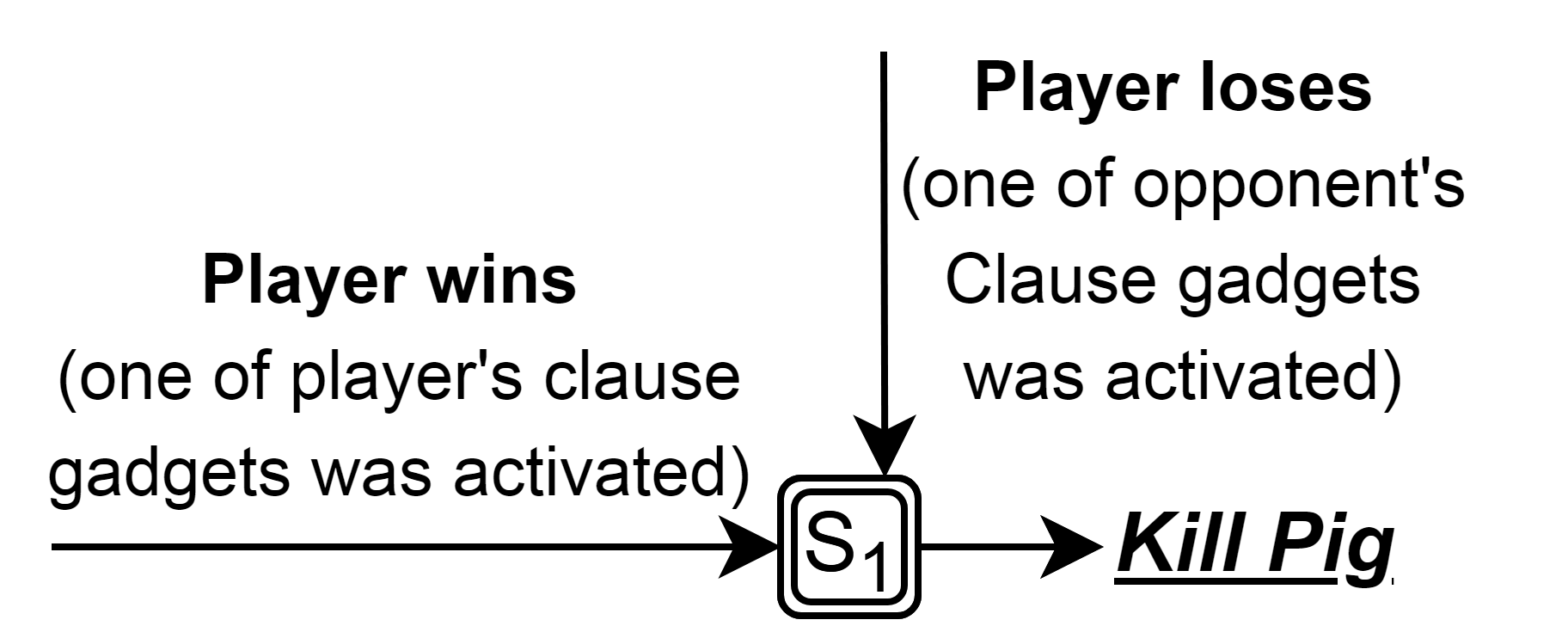}
  \caption{Model of the Result gadget used.}
\end{figure}

\subsubsection{Level Construction}
Now that all the necessary gadgets have been described, the only remaining requirement is that they can be successfully arranged throughout the level space.

\begin{lemmaz}
Any given game of G2 can be reduced to an ABES level definition in polynomial time.
\end{lemmaz}

\begin{zproof}
As we have already shown that each of the necessary gadgets can be created using a polynomial amount of space and elements and can therefore also be described in polynomial time, the only remaining requirement is that all the gadgets can be successfully arranged throughout the level in polynomial time, relative to the size of the G2 setup description (two 12-DNF Boolean formulas, variable assignment and initial variable values). As the number of gadgets required is clearly polynomial, it suffices to describe a polynomial time method for determining the location of each gadget, as well as the level's width, height, slingshot position and number of birds.

By using the same reasoning as in our PSPACE-hard level construction (Lemma 4.14), we know that the time required to compute the relative placement (spatial arrangement) of these gadgets, as well as the space between them, is polynomial in the total number of gadgets. There are also always a polynomial number of tunnels between gadgets and each tunnel can always be connected to its appropriate destination in polynomial time. Because of this, we can be certain than an equivalent ABES level description for any given game of G2 can always be created in a polynomial amount of space relative to the length of the original Boolean formulas, and thus it can also be defined in polynomial time. All calculations for slingshot position, release points needed, level's width/height, etc., can be calculated the same as in Section 4.3.

Lastly, the number of birds the player has is equal to $(2V_{p}+4)(2^{V_{N}})$, where $V_{p}$ is equal to the total number of variables assigned to the player, and $V_{N}$ is equal to the total number of variables assigned to both the player and the opponent. This is equivalent to the maximum number of birds required to make a move for both the player and opponent (four birds needed for the Ordering gadget, as well as $2V_{p}$ possible literal options in the Choice gadget), multiplied by the maximum number of possible value combinations for all variables $(2^{V_{N}})$. If the player cannot win the level in this many birds, then at least one of the variable value combinations has been repeated. 
\end{zproof}

An example diagram of a fully constructed structure, using the same Boolean formula as in Figure 16, is shown in the Appendix (Figure A.30). For this example, the player is assigned the variables $z$ and $w$, the opponent is assigned the variables $x$ and $y$, and all variables are initially given a negative truth value.  

As we have constructed the necessary gadgets and can position them within the game's environment in polynomial time, the problem of solving levels for ABES is EXPTIME-hard.

\begin{theorem}
The problem of solving levels for ABES is EXPTIME-hard.
\end{theorem}

\subsection{Winning Strategy (Example)}
We now describe an example of a winning strategy for solving an ABES level description that has been reduced from the Boolean formulas for the player and opponent given in Figure 16. For this example, the player is assigned the variables $z$ and $w$, the opponent is assigned the variables $x$ and $y$, and all variables are initially given a negative truth value (same setup as for the example structure diagram in Figure A.30). 
For this level description, we can see that the player will immediately need to set the value of $w$ to positive. If the player doesn't do this then there is a chance that variable $y$ would be changed to positive when the opponent makes their move, which would mean that the opponent's second clause would be satisfied (leading to a loss). To set the variable $w$ to positive we need to open all AUT gates in the Choice gadget except for the last one. We can then traverse the AUT gates in the Choice gadget via the Ordering gadget, which will subsequently adjust the Clause gadgets to represent $w$ now being positive. We then need to make a random move for the opponent, and check if any of their associated Clause gadgets are activated (none of them are regardless of the outcome of the opponent's random move). After this, we should see that we only need to set the value of the variable $z$ to positive to satisfy one of our clauses. This is the case regardless of what move was previously made for the opponent, although the specific clause that is satisfied might change. After setting $z$ to positive we can then check our clauses for satisfiability, and as one of our Clause gadgets is activated a bird will pass through the Result gadget and solve the level.

\section{Proof Generalisation}
The complexity proofs described in this paper can be replicated in many other games similar to Angry Birds, as long as the necessary gadgets can be constructed. In general, this means that the computational complexity of any physics-based game can be established using our frameworks, as long as the following requirements hold. A level within the game contains a set number of targets, which the player needs to hit or reach in order to solve the level. The game contains both static and non-static elements. The game contains elements that can be moved as a result of the player's actions. The physics engine utilised by the game allows for rudimentary systems of gravity, momentum, energy transfer and rotational motion (almost all simple physics engines should contain this). The player cannot directly influence any element within a gadget framework, instead only being able to interact with it through the use of a secondary non-static game element (in our case a bird), which enters the gadget framework through designated entry points. No new element can enter this framework until the outcome of any previously entered element is finalised. For our EXPTIME-hardness proof, we also require the exact outcome of certain player actions to be unknown beforehand.

Whilst by no means applicable to all games that contain these features, this generalisation suggests that many other physics-based games are NP-hard and/or PSPACE-complete. This includes both games that are similar in play style to Angry Birds, such as Crush the Castle, Siege Hero or Fragger, as well as games that play considerably differently, such as Where's My Water, World of Goo, Bad Piggies, Cut the Rope 2, Crayon Physics Deluxe, The Incredible Machine, Eets and Peggle, to name just a few. Even though formal proofs on the complexity of these games would likely each be as long as this paper again, we provide below some rough outlines for how single-use EQ and Clause gadgets could be implemented for several popular examples of other physics-based games. Single-use EQ gadgets can only be used to set the value of their associated variable once, while single-use Clause gadgets remain activated once they are activated the first time (i.e. can't be un-activated). While these single-use gadgets are much less sophisticated than those we presented previously, they can still be used for NP-hardness proofs based on our 3-SAT reduction framework as only a single framework cycle is needed.

\subsection{Where's My Water}
The aim of this game is to get a certain number of water droplets into a specific destination pipe. These water droplets behave in the same manner as red birds in Angry Birds, after they have been fired from the slingshot. The game contains dirt areas that water droplets cannot pass through, but which the player can remove by tapping them. The game also contains doors that stop water droplets when closed. Each door has a button associated with it. When the button associated with a door is pressed by a water droplet, the door opens. The game also contains pipes that allow water droplets to pass each other without any risk of leakage or collision. An example level from Where's My Water \cite{water} is shown in Figure 24.

\textbf{EQ gadget:} Each EQ gadget contains a single water droplet and two possible tunnels on either side of it that are blocked by dirt. The player can remove this dirt by tapping on it, allowing them to direct the water droplet into either tunnel. Whichever tunnel the player directs the water droplet into indicates the value to set the associated variable to (i.e. if the water droplet falls into the left/right tunnel then set the value of the variable to negative/positive). As there is only one water droplet in each EQ gadget, the player can only set the value of the associated variable once (i.e. this EQ gadget is single-use only).

\textbf{Clause gadget:} Each Clause gadget contains a button that, when touched by a water droplet, opens a door that releases a set number of water droplets into the destination pipe. When the player indicates the truth value for a variable using its associated EQ gadget, the water droplet will travel through all the Clause gadgets that contain the chosen literal, pressing the button within any Clause gadget it travels through (i.e. pressing the button within a Clause gadget will essentially activate it). As the effect of pressing the button within a Clause gadget cannot be undone, these Clause gadgets are single-use only.

\textbf{Crossover:} Pipes can simply be used to allow water droplets to travel over one another.  

\textbf{Victory condition:} The level is solved once all Clause gadgets have released their water droplets into this destination pipe (i.e. when all Clause gadgets are activated).

\subsection{Cut the Rope 2}
The aim of this game is to transport a piece of candy to a stationary creature. This piece of candy behaves in the same manner as red birds in Angry Birds, after they have been fired from the slingshot. The game contains balloons which can hold objects in a specific place (i.e. the object becomes unaffected by gravity). If an object is connected to one balloon then it is suspended a fixed distance directly below this balloon. If an object is connected to several balloons then it is suspended a fixed distance below the mid-point between these balloons. The player can remove a balloon by tapping on it (i.e. ``pop'' the balloon). The game also contains wooden balls that behave the same as the piece of candy. The game also contains rotating doors (gear attached to a wooden block) that objects cannot pass through when closed. Each door has a button associated with it. When the button associated with a door is pressed by an object, the door opens. An example level from Cut the Rope 2 \cite{rope} is shown in Figure 25.

\textbf{EQ gadget:} Each EQ gadget contains a wooden ball that is suspended in place by two balloons, and two possible tunnels on either side of the wooden ball. The player can pop each of these balloons by tapping them. The order in which the two balloons suspending the wooden ball are popped can be used to direct the wooden ball into either tunnel. Whichever tunnel the player directs the wooden ball into indicates the value to set the associated variable to. As there is only one wooden ball in each EQ gadget, the player can only set the value of the associated variable once.

\textbf{Clause gadget:} Each Clause gadget contains a button that when touched by a wooden ball, opens a rotating door outside of the framework. When the player indicates the truth value for a variable using its associated EQ gadget, the wooden ball will travel through all the Clause gadgets that contain the chosen literal, pressing the button within any Clause gadget it travels through (i.e. activates the Clause gadget).

\textbf{Crossover:} Crossover gates can be constructed using the exact same design as for Angry Birds (Section 3.4).

\textbf{Victory condition:} The piece of candy is suspended by a balloon above a stack of rotating doors placed outside the rest of the framework. Each rotating door in this stack is turned on when one of the Clause gadgets is activated (i.e. each button in a Clause gadget turns on one of these rotating doors). The creature is placed below this stack of rotating doors. The player can pop the balloon suspending the piece of candy at any point, but the candy can only reach the creature (i.e. solve the level) if all rotating doors are turned on (i.e. if all Clause gadgets are activated).

\begin{figure}[t]
\centering
\begin{minipage}{.49\linewidth}
  \centering
  \includegraphics[width=0.9\linewidth]{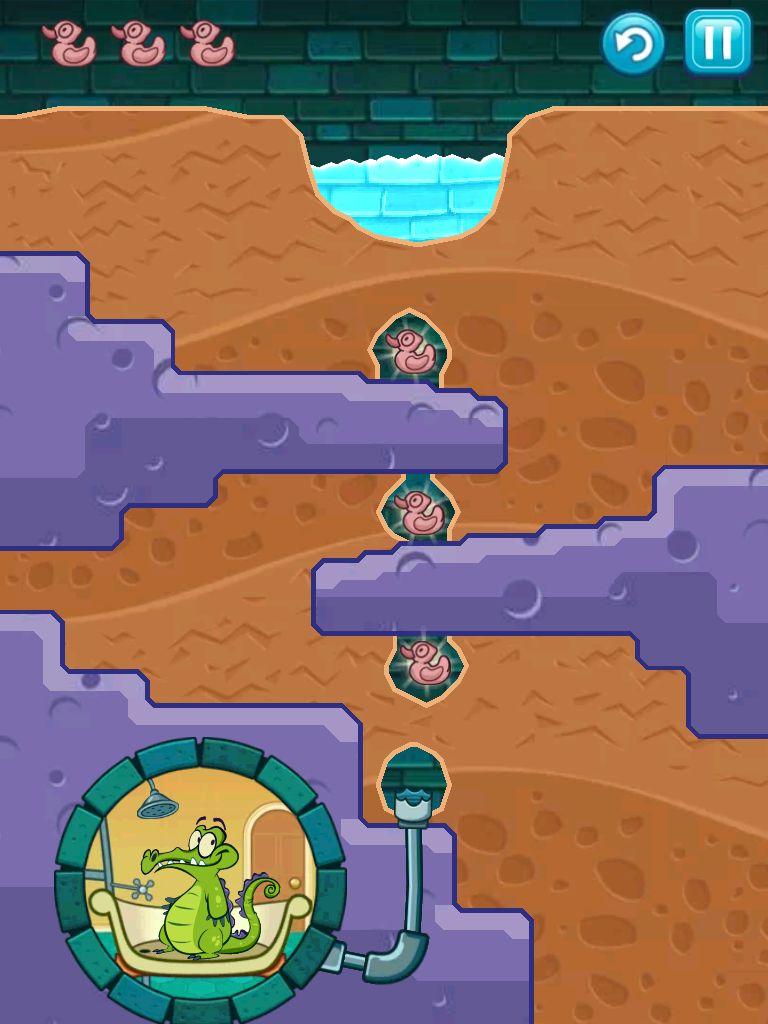}
\captionsetup{justification=centering}
  \captionof{figure}{Screenshot of a level for Where's My Water.}
  \label{fig:test2}
\end{minipage}
~
\begin{minipage}{.49\linewidth}
  \centering
  \includegraphics[width=.9\linewidth]{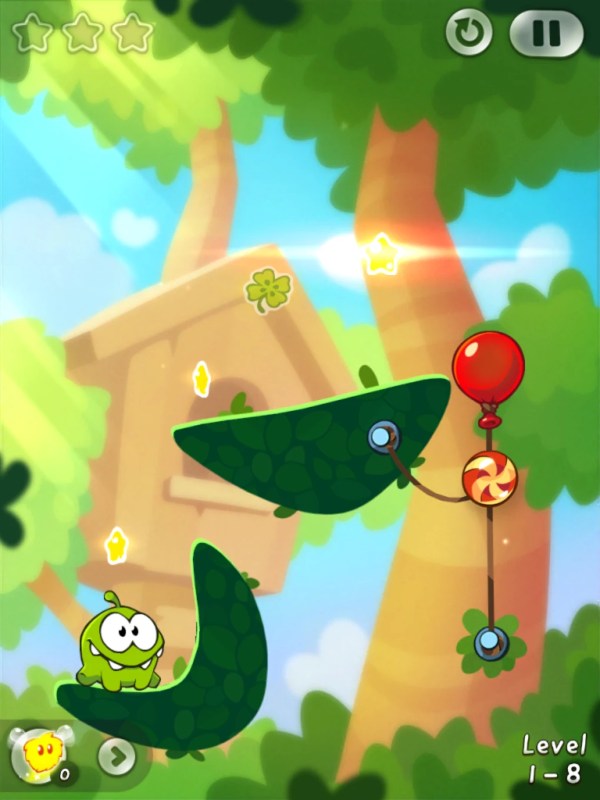}
\captionsetup{justification=centering}
  \captionof{figure}{Screenshot of a level for Cut the Rope 2.}
  \label{fig:test1}
\end{minipage}
\end{figure}

\subsection{The Incredible Machine}
The aim of this game is to accomplish some predefined task for a given environment by placing objects within the level. For our setup, the only objects that the player can place in the level are candles. The game contains baseballs that behave in the same manner as red birds in Angry Birds, after they have been fired from the slingshot. The game also contains brick walls that objects cannot pass through, and TNT that can be ignited with candle. When a TNT is ignited it will explode and destroy (remove) both itself and any objects (such as walls) next to it. The game also contains torches that can be turned on by an object hitting them. The game also contains pipes that allow objects to pass each other without any risk of leakage or collision. An example level from The Incredible Machine \cite{machine} is shown in Figure 26.

\textbf{EQ gadget:} Each EQ gadget contains a baseball and two possible tunnels on either side of it that are blocked by brick walls. TNT is placed next to each of these brick walls and can be ignited by placing a candle next to it. When a TNT is ignited it will explode and destroy both itself and the brick wall next to it. Igniting one of these TNTs can therefore be used to direct the baseball into either tunnel. Whichever tunnel the player directs the baseball into indicates the value to set the associated variable to. As there is only one baseball in each EQ gadget, the player can only set the value of the associated variable once.

\textbf{Clause gadget:} Each Clause gadget contains a torch. When the player indicates the truth value for a variable using its associated EQ gadget, the baseball will travel through all the Clause gadgets that contain the chosen literal, hitting and turning on the torch within any Clause gadget it travels through (i.e. activates the Clause gadget).

\textbf{Crossover:} Pipes can simply be used to allow baseballs to travel over one another.  

\textbf{Victory condition:} The requirement for solving the level is set to turning on all of the torches within the Clause gadgets (i.e. when all Clause gadgets are activated).

\vskip 0.1in

While proofs for NP-hardness and PSPACE-hardness can often be generalised between different video games, our proposed proof of EXPTIME-hardness is trickier to replicate. We postulate though that it might be possible to prove that extended versions of other popular games such as Super Mario Bros. are EXPTIME-hard by introducing elements such as ``mystery'' boxes which could spawn a random item, thus providing the necessary uncertainty in player actions. However, a more thorough investigation and research would be needed to determine if this is possible.

\begin{figure}[t]
  \centering
  \includegraphics[width=0.7\linewidth]{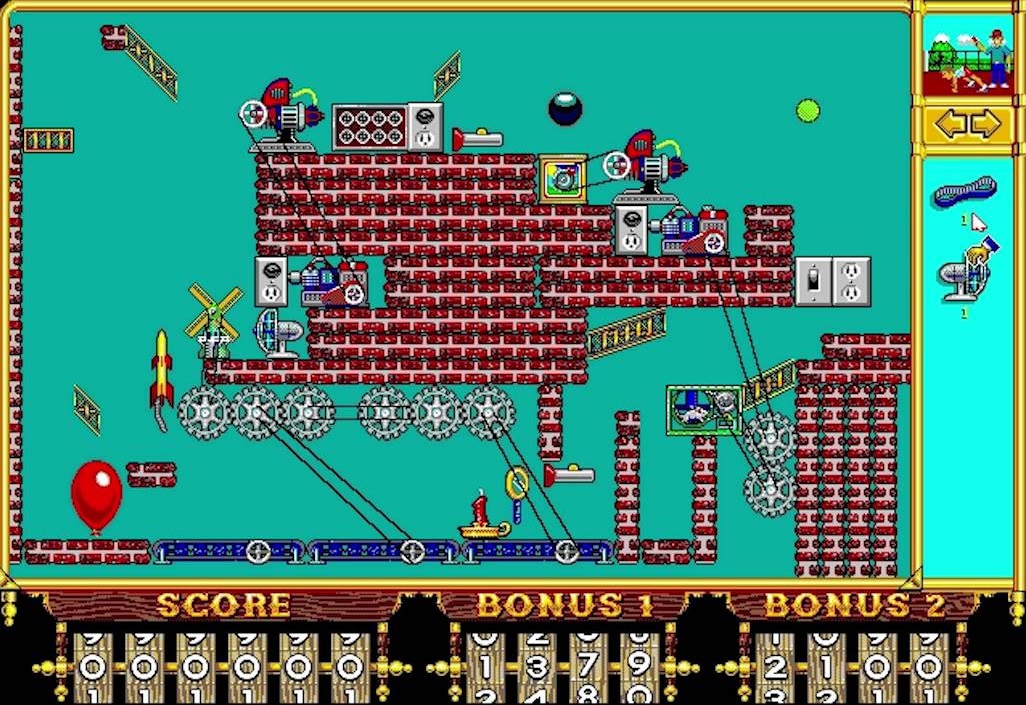}
  \caption{Screenshot of a level for The Incredible Machine.}
\end{figure}

\section{Conclusions}
In this paper, we have proven that the task of deciding whether a given Angry Birds level can be solved is either NP-hard, PSPACE-hard, PSPACE-complete or EXPTIME-hard, depending on the version of the game being used. 

To the best of our knowledge, this is the first example of a single-player game without a traditional opponent being proved EXPTIME-hard. Our use of unknown and changing environmental variables as the opponent which the player is facing, is a unique view of the problem and opens up the possibility of proving many other games EXPTIME-hard using this methodology. The most likely candidates for this analysis would be games that feature some inherent stochasticity in their engine (similar to the method employed for our proof), or games which use randomness within one of their gameplay elements (such as mystery/question blocks in Mario games). In games like this the player may know what elements the box could contain, but will not know exactly what it does contain until after they open it. This would be a good basis for constructing an opponent for a reduction from G2 or another similar EXPTIME-complete game. It is also possible to use the inaccuracy of the player's input or another similar area of uncertainty to generate the required randomness. EXPTIME-hardness proofs might also be able to be applied to real-world environments.

This work provides an important contribution to the collection of games that have been investigated within the field of computational complexity. However, there is still a huge assortment of physics-based and other non-traditional puzzle games that are available for future analysis, which do not follow the typical structure of those previously studied. The importance of games for AI research lies in the fact that games can form a simplified and controlled environment, which allows for the development and testing of AI methods that will eventually be used in the real world. It is also highly likely that the proofs presented in this paper can be generalised to other physical reasoning and AI problems. Even though Angry Birds may initially seem like a simple game, the challenges that dealing with its physics simulation engine pose make it incredibly relevant to those in the real world. We are therefore hopeful that this work will inspire future research into a more diverse range of game types and problems.

\section*{Acknowledgments}
We would like to thank the three reviewers for their incredibly detailed reviews and the many excellent suggestions they made for improving this paper.

\bibliography{mybibfile}

\par\vspace*{\fill}
\appendix
\section{Full structure construction examples (not to scale)}

\begin{figure}
  \centering
  \includegraphics[width=0.92\linewidth]{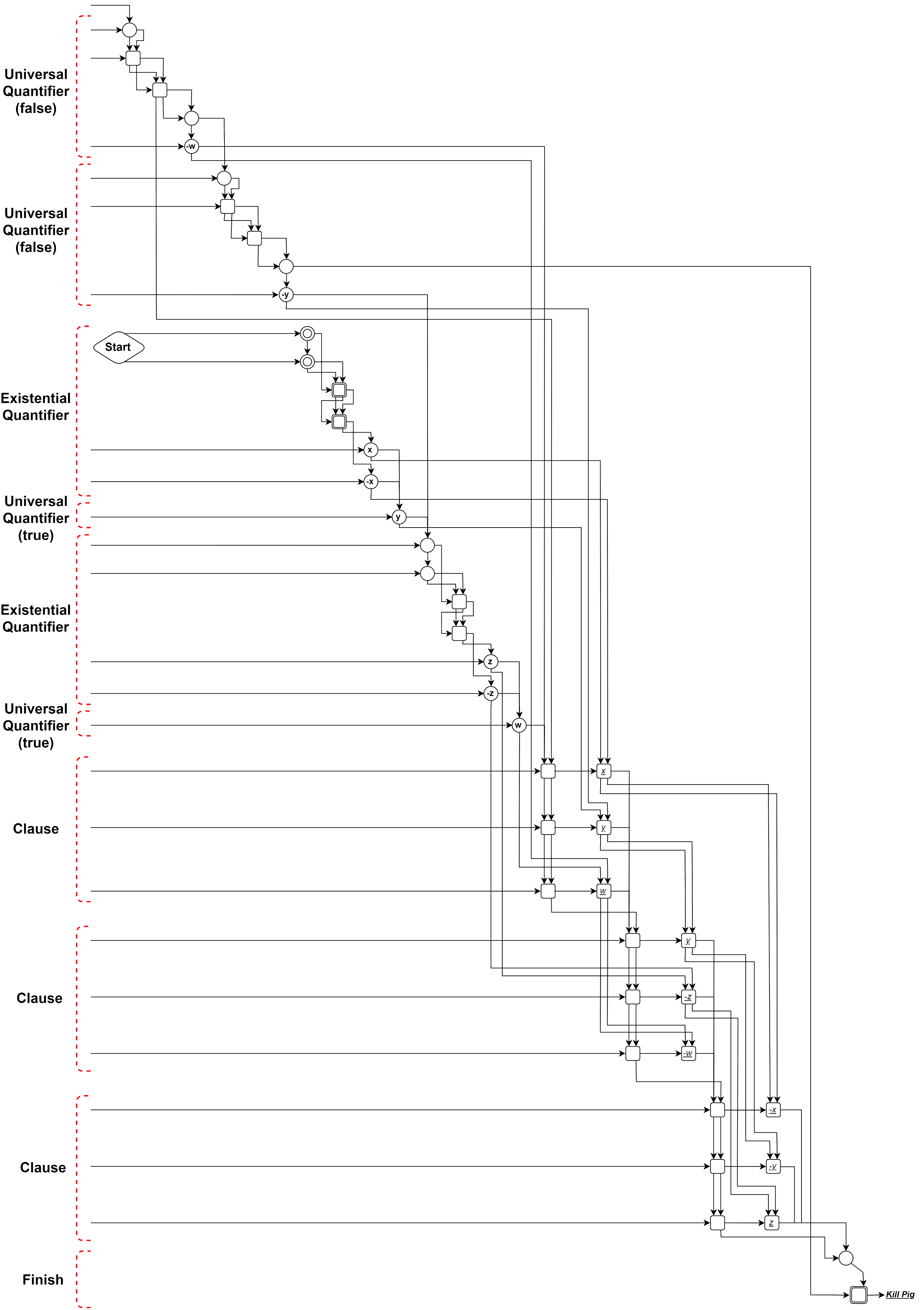}
  \caption{ABED (PSPACE-complete)}
\end{figure}

\begin{figure}
  \centering
  \includegraphics[width=0.92\linewidth]{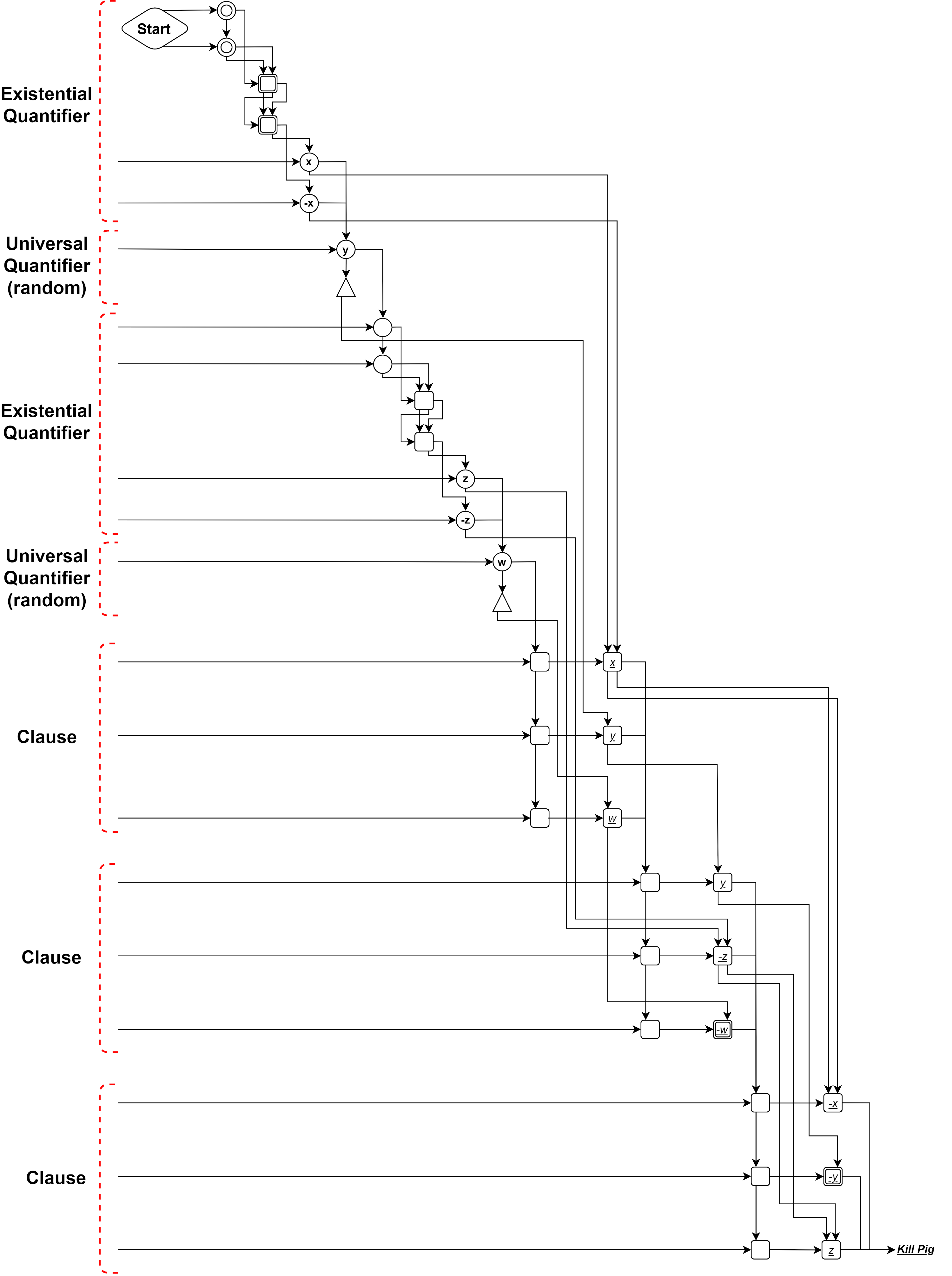}
  \caption{ABPS (PSPACE-hard)}
\end{figure}

\begin{figure}
  \centering
  \includegraphics[width=0.92\linewidth]{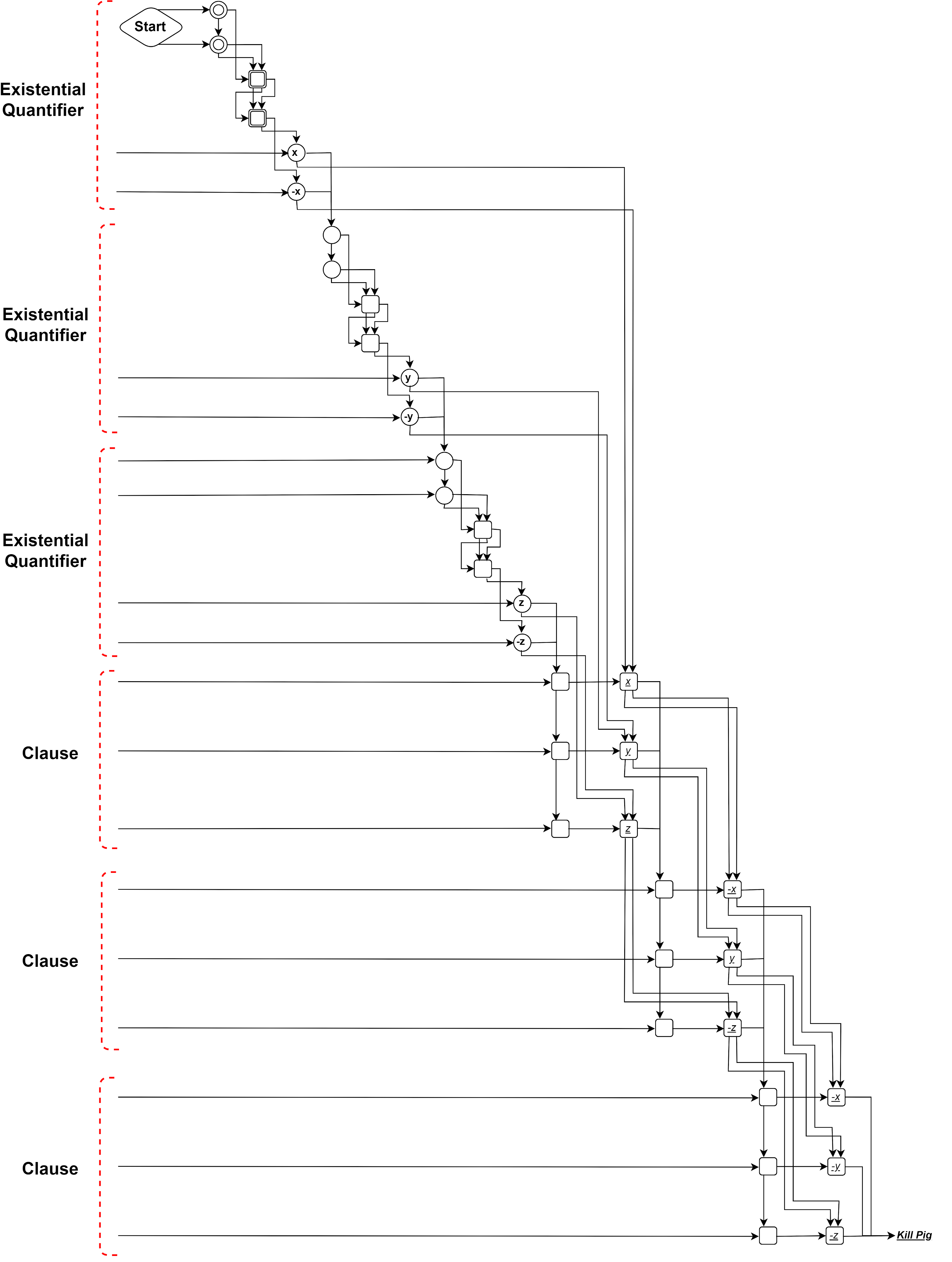}
  \caption{ABPD (NP-hard)}
\end{figure}

\begin{figure}
  \centering
  \includegraphics[width=0.82\linewidth]{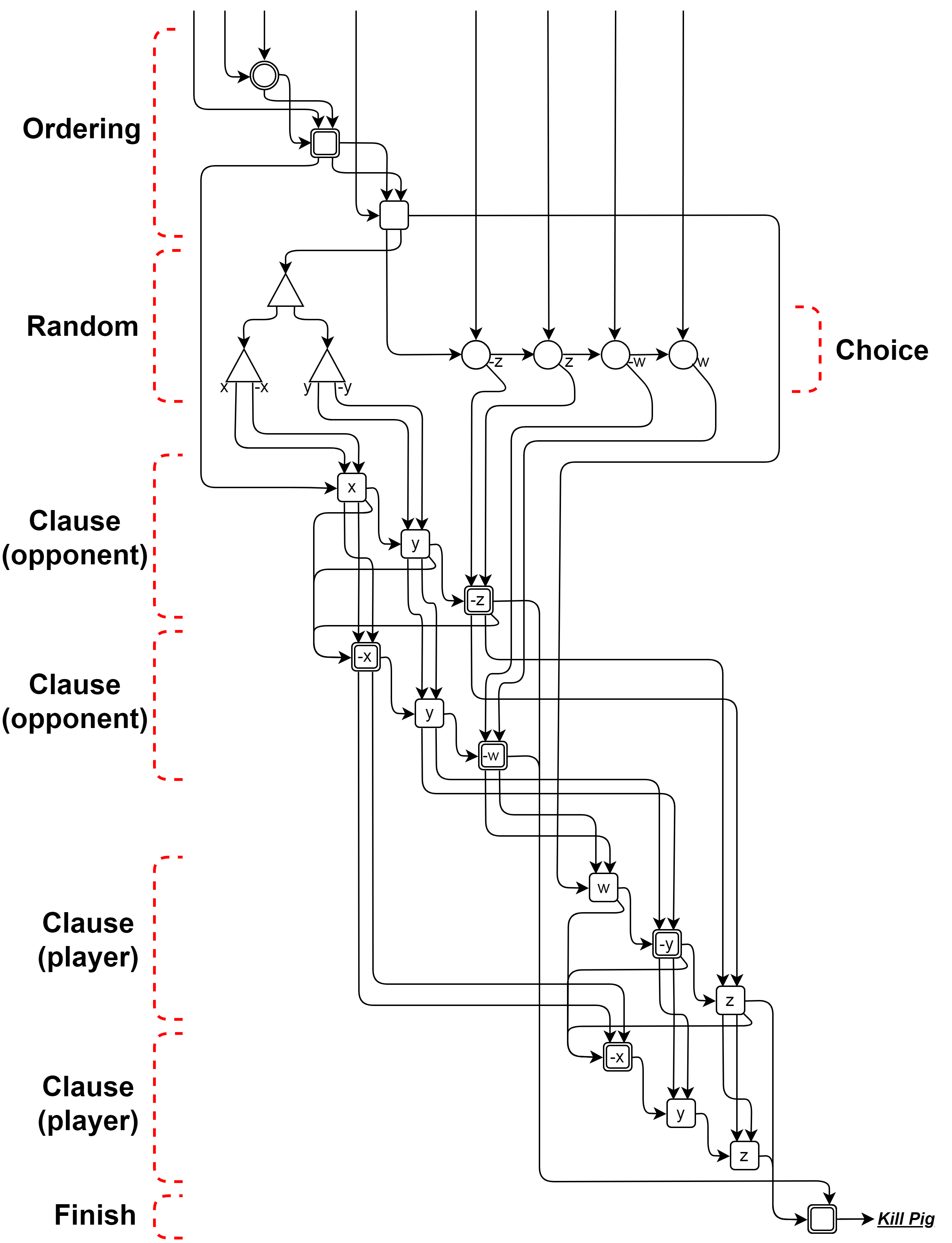}
  \caption{ABES (EXPTIME-hard)}
\end{figure}

\newpage

\section{Step-by-step shot ordering}

\begin{figure}[h]
  \centering
  \includegraphics[width=0.99\linewidth]{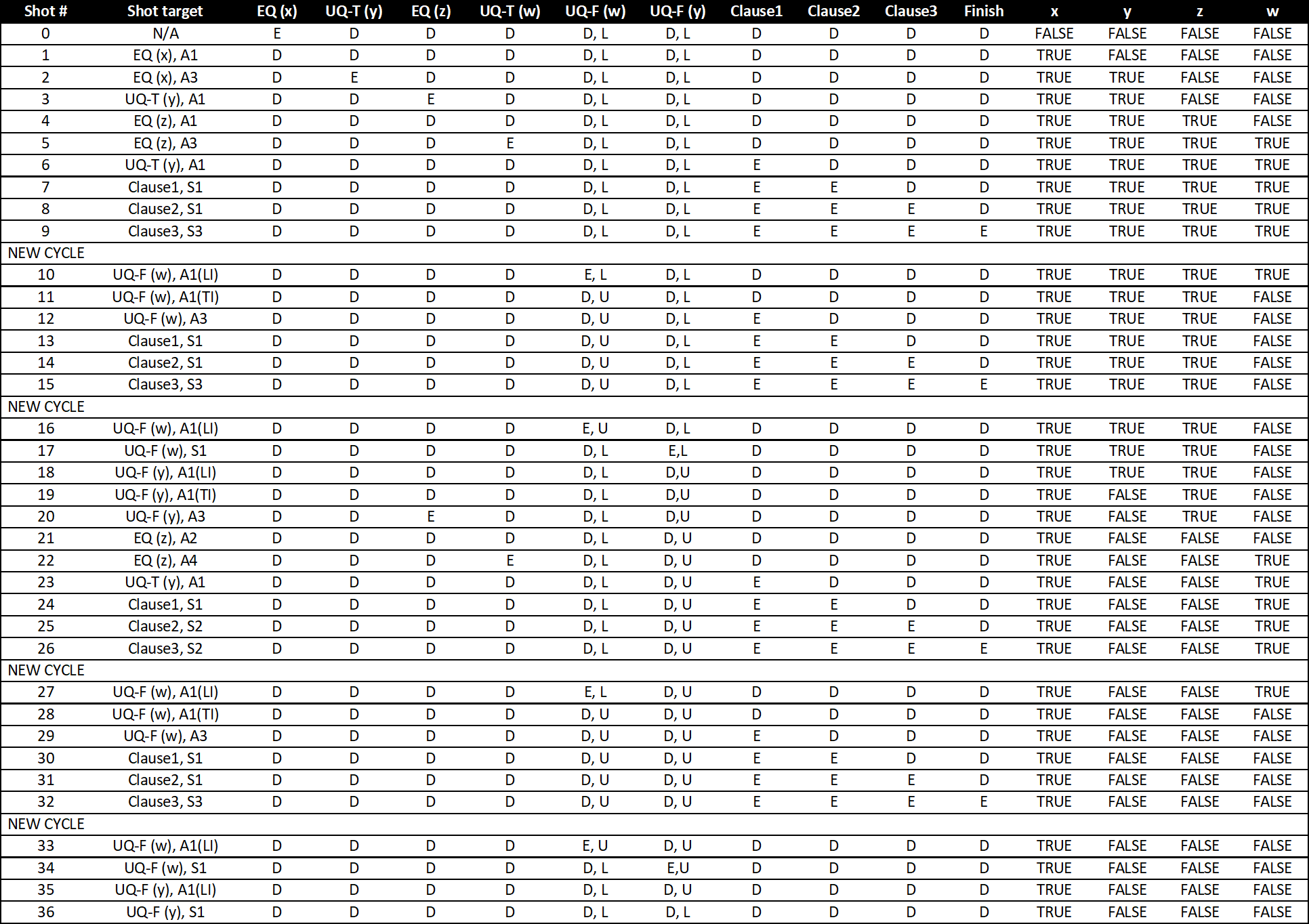}
  \caption{Shots required to solve the level created using the example ABED framework shown in Figures 7 and A.24. Each row of the table specifies the target for each shot (gadget, gate, and entrance tunnel when ambiguous), the state of each gadget after the shot has resolved (whether it is enabled (E), disabled (D), locked (L), or unlocked (U)), and the truth value of each Boolean variable.}
\end{figure}

\par\vspace*{\fill}
\section{Gadget truth tables}

This appendix provides detailed truth tables for each gadget described in this paper (except for the trivial cases). Empty cells for the current state indicate that the gate in question can be either open or closed. The bird input point specifies the gate by which the bird entered the gadget, as well as the specific gate entrance point when ambiguous. Bird input points with the ``(Enable)'' marker represent that the gadget is enabled by a bird entering here. Empty cells for the next state indicate that the position of the gate in question is unchanged. Empty cells for the Output indicate that the bird did not exit the gadget. 

\begin{figure}[h]
  \centering
  \includegraphics[width=0.99\linewidth]{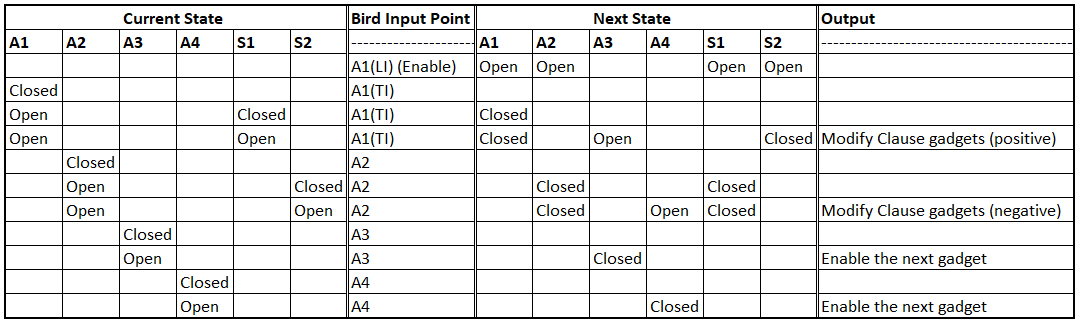}
  \caption{EQ gadget truth table.}
\end{figure}

\begin{figure}[h]
  \centering
  \includegraphics[width=0.55\linewidth]{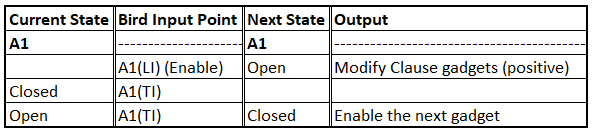}
  \caption{UQ-T gadget truth table.}
\end{figure}

\begin{figure}[h]
  \centering
  \includegraphics[width=0.9\linewidth]{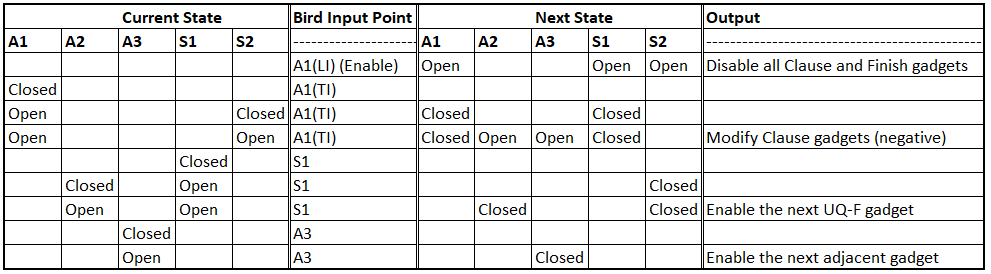}
  \caption{UQ-F gadget truth table.}
\end{figure}

\begin{figure}[h]
  \centering
  \includegraphics[width=0.99\linewidth]{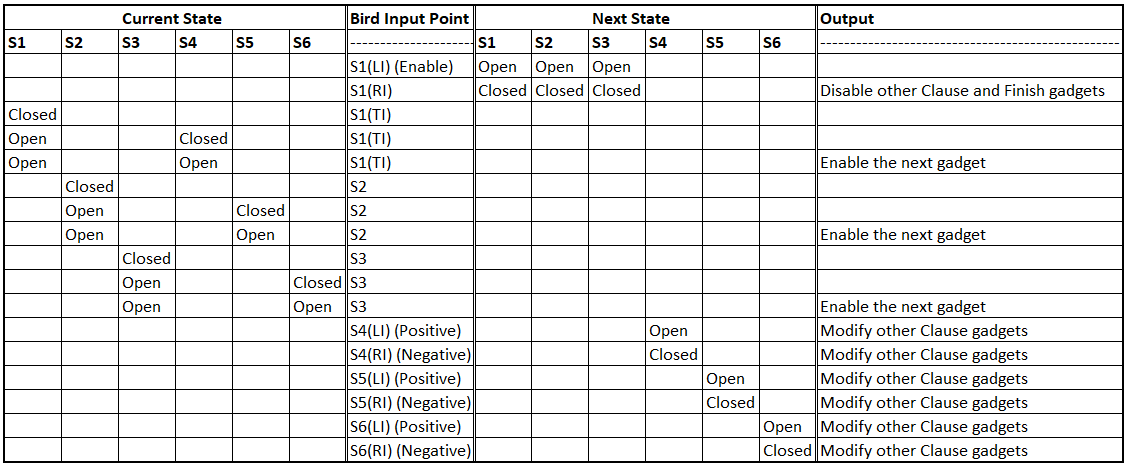}
  \caption{Clause gadget truth table (ABED).}
\end{figure}

\begin{figure}[h]
  \centering
  \includegraphics[width=0.6\linewidth]{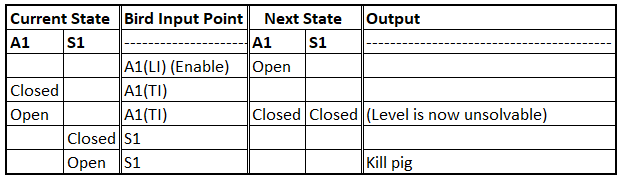}
  \caption{Finish gadget truth table.}
\end{figure}

\begin{figure}[h]
  \centering
  \includegraphics[width=0.75\linewidth]{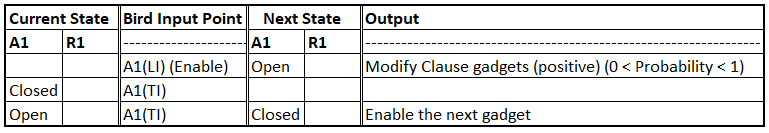}
  \caption{UQ-R gadget truth table.}
\end{figure}

\begin{figure}[h]
  \centering
  \includegraphics[width=0.65\linewidth]{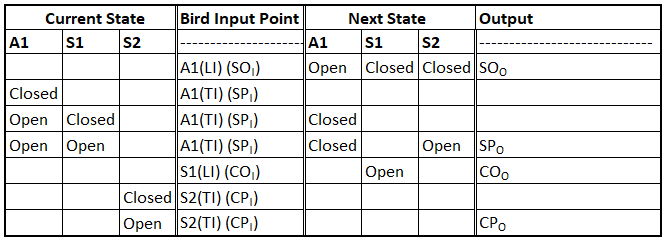}
  \caption{Ordering gadget truth table.}
\end{figure}

\begin{figure}[h]
  \centering
  \includegraphics[width=0.7\linewidth]{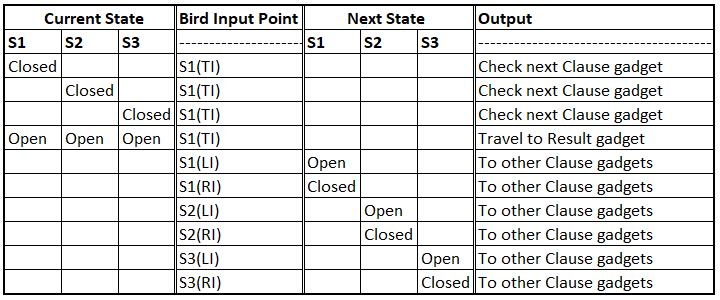}
  \caption{Clause gadget truth table (ABES) example for a clause with three literals.}
\end{figure}

\end{document}